\documentclass[10pt]{revtex4}
\usepackage{latexsym}
\usepackage{epsfig}
\usepackage{amsmath}
\begin{document}
\title{A note on holographic dark energy with varying $c^2$ term}
\author{A. Sheykhi$^{1,2}$,\footnote{
Email:asheykhi@shirazu.ac.ir} S. Ghaffari$^2$ and N.
Roshanshah$^1$}
\address{$^1$  Physics Department and Biruni Observatory, College of
Sciences, Shiraz University, Shiraz 71454, Iran\\
$^2$Research Institute for Astronomy and Astrophysics of Maragha
(RIAAM), P.O. Box 55134-441, Maragha, Iran}

\begin{abstract}
We reconsider the holographic dark energy (HDE) model with a
slowly time varying $ c^2(z)$ parameter in the energy density,
namely $\rho_D=3M_p^2 c^2(z)/L^2$, where $L$ is the IR cutoff and
$z$ is the redshift parameter. As the system's IR cutoff we choose
the Hubble radius and the Granda-Oliveros (GO) cutoffs. The latter
inspired by the Ricci scalar curvature. We derive the evolution of
the cosmological parameters such as the equation of state and the
deceleration parameters as the explicit functions of the redshift
parameter $z$. Then, we plot the evolutions of these cosmological
parameters in terms of the redshift parameter during the history
of the universe. Interestingly enough, we observe that by choosing
$L=H^{-1}$ as the IR cutoff for the HDE with time varying $ c
^2(z) $ term, the present acceleration of the universe expansion
can be achieved, even in the absence of interaction between dark
energy and dark matter. This is in contrast to the usual HDE model
with constant $c^2$ term, which leads to a wrong equation of
state, namely that for dust $w_D=0$, when the IR cutoff is chosen
the Hubble radius.
\end{abstract}

\maketitle

\section{Introduction}
According to nowadays observations, present acceleration of the
universe expansion has been well established \cite{Riess}. Within
the framework of general relativity, the responsible component of
energy for this accelerated expansion is known as \textit{dark
energy} (DE) with negative pressure. However, the nature of  DE is
still unknown, and some candidates have been proposed to explain
it. The earliest and simplest candidate is the cosmological
constant with the time independent equation of state $
\omega_\Lambda=-1 $ which has some problems like fine-tuning and
coincidence problems. Therefore, other theories have been
suggested for the dynamical DE scenario to describe the
accelerating universe.

An interesting attempt for probing the nature of DE within the
framework of quantum gravity, is the so-called HDE proposal. This
model which has arisen a lot of enthusiasm recently
\cite{Coh,Li,Huang,Hsu,HDE,Setare1,wang0,SheyHDE}, is motivated
from the holographic hypothesis \cite{Suss1} and has been tested
and constrained by various astronomical observations \cite{Xin}.
In holographic principle a short distance  cutoff could be related
to a long distance cutoff (infrared cutoff) due to the limit set
by formation of a black hole.  Based on the holographic principle,
it was shown by Cohen et al. \cite{Coh} that the quantum
zero-point energy of a system with size $L$ should not exceed the
mass of a black hole with the same size, i.e.,
\begin{equation}
L^3\rho_D\leq L M_p^2,
\end{equation}
where $M^2_p = 8\pi G$ is the reduced Planck mass and $L$ is the
IR cutoff. The largest $L$ allowed is the one saturating  this
inequality so that we get the
\begin{equation}\label{HDE}
\rho_D=3c^2M_p^2/L^{2},
\end{equation}
where $ 3 c^2 $ is a dimensionless constant. There are many models
of HDE, depending on the IR cutoff, that have been studied in the
literatures \cite{Horava,Fischler,Nojiri,Gao}. The simple choice
for IR cutoff is the Hubble radius, i.e., $ L = H^{-1}$ which
leads to a wrong equation of state (EoS) and the accelerated
expansion of the universe cannot be achieved. However, as soon as
an interaction between HDE and dark matter is taken into account,
the identification of IR cutoff with Hubble radius $H^{-1}$, in
flat universe, can simultaneously drive accelerated expansion and
solve the coincidence problem. \cite{Pavon}. Then, Li \cite{Li}
showed that taking the particle horizon radius as IR cutoff it is
impossible to obtain an accelerated expansion. He also
demonstrated that the identification of $L$ with the radius of the
future event horizon gives the desired result, namely a
sufficiently negative equation of state to obtain an accelerated
universe.

It is worth noting that, for the sake of simplicity, very often
the $c^2$ parameter in the HDE model is assumed constant. However,
there are no strong evidences to demonstrate that $c^2$ should be
a constant and one should bear in mind that it is more general to
consider it a slowly varying function of time. It has been shown
that the parameter $c^2$ can play an essential role in
characterizing the model. For example, it was argued that the HDE
model in the far future can be like a phantom or quintessence DE
model depending whether the parameter $c^2$ is larger or smaller
than $1$, respectively \cite{Radicella}. By slowly vary function
with time, we mean that $(\dot{c}^2)/(c^2)$ is upper bounded by
the Hubble expansion rate, i.e., \cite{Radicella}
\begin{equation}
\frac{({c}^2)^{\dot{}}}{c^2}\leq H,
\end{equation}
where dot indicates derivative with respect to the cosmic time. In
this case the time scale of the evolution of $c^2$ is shorter than
$H^{-1}$ and one can be satisfied to consider the time dependency
of $c^2$ \cite{Radicella}. Considering the future event horizon as
IR cutoff, the HDE model with time varying parameter $c^2$, has
been studied in  \cite{Malekjani}. It was argued that depending on
the parameter $c^2$, the phantom regime can be achieved earlier or
later compared  to the usual HDE  with constant $c^2$ term
\cite{Malekjani}. In this paper, we reconsider the HDE model with
the slowly varying parameter $c^2(z) $ by taking into account the
Hubble horizon $L=H^{-1}$ and GO cutoff, $L=(\alpha
H^2+\beta\dot{H})^{{-1}/{2}}$, as the system's IR cutoffs. We
shall study four parameterizations of $c(z)$ as follows
\begin{eqnarray}
&&\text{GHDE1}:~~~~c(z)=c_0+c_1\frac{z}{(1+z)},\label{G1} \\
&&\text{GHDE2}:~~~~ c(z)=c_0+c_1\frac{z}{(1+z)^2},\label{G2} \\
&&\text{GHDE3}:~~~~ c(z)=\frac{c_0}{1+c_1\ln(1+z)},\label{G3}\\
&&\text{GHDE4}:~~~~c(z)=c_0+c_1\left(\frac{\ln(2+z)}{1+z}-\ln2\right),
\label{G4}
\end{eqnarray}
where  GHDE  stands  for the Generalized HDE  model. The above
choices for $c(z)$ are, respectively, inspired by the
parameterizations known as Chevallier-Polarski-Linder
parametrization (CPL) \cite{Chevallier}, Jassal-Bagla-Padmanabhan
(JBP) parametrization \cite{Jassal}, Wetterich parametrization
\cite{Wetterich}, and Ma-Zhang parametrization \cite{Ma}. Setting
$ c_1=0 $ in all these four parameterizations, the original HDE
with constant $c$ parameter is recovered.

This paper is organized as follows. In section II, we drive the
basic equations for the HDE with time varying $c^2$  parameter. In
this section, we also consider the Hubble radius as IR  cutoff and
derive the evolution of EoS and deceleration parameters by
choosing $c(z)$. In section III, we repeat the study for the GO
cutoff and investigate the evolution of the cosmological
parameters. The last section is devoted to conclusions and
discussions.
\section{GHDE in flat FRW cosmology with Hubble radius as IR cutoff}
In the context of flat Friedmann-Robertson-Walker (FRW) cosmology,
the Friedmann equation can be written
\begin{equation}\label{Friedeq1}
H^2=\frac{1}{3M_p^2}(\rho_m+\rho_D),
\end{equation}
where $\rho_m$ and $ \rho_D $ are the energy densities of
pressureless dark matter and DE, respectively. By using the
dimensionless energy densities
\begin{equation}\label{Omega}
\Omega_m=\frac{\rho_m}{3m_p^2H^2},~~~~~~~~~~~~~~~~~~\Omega_D=\frac{\rho_D}{3m_p^2H^2},
\end{equation}
the Friedmann equation (\ref{Friedeq1}) can be written as
\begin{equation}\label{Friedeq2}
\Omega_m+\Omega_D=1.
\end{equation}
We shall  assume there is no interaction between dark matter and
GHDE. Therefore, both components obey independent conservation
equation.  The conservation equations for pressureless dark matter
and DE, are given by
\begin{eqnarray}\label{ConserveCDM}
&&\dot{\rho}_m+3H\rho_m=0,\\
&&\dot{\rho}_D+3H(1+\omega_D)\rho_D=0,\label{ConserveDE}
\end{eqnarray}
where $w_D=p_D/ \rho_D$ is the EoS parameter of GHDE. In this
section we consider the Hubble radius as IR cutoff, $ (L=H^{-1})
$, thus the energy density of GHDE model from (\ref{HDE}) can be
written as
\begin{equation}\label{GHDE}
\rho_D=3M_p^2c^2(z)H^2.
\end{equation}
Using definition (\ref{Omega}), the dimensionless energy density
for the GHDE becomes
\begin{equation}\label{GHDE1}
\Omega_D(z)=c^2(z).
\end{equation}
Taking the time derivative of Eq. (\ref{GHDE}), we find
\begin{equation}\label{rhodot}
\dot{\rho}_D=2\rho_D\left(\frac{\dot{c}(z)}{c(z)}+\frac{\dot{H}}{H}\right).
\end{equation}
Besides, if we take the time derivative of Friedmann equation
(\ref{Friedeq1}), after using Eqs. (\ref{Friedeq2}),
(\ref{ConserveCDM}) and (\ref{rhodot}), we find
\begin{equation}\label{Hdot1}
\frac{\dot{H}}{H^2}=-\frac{3}{2}+\frac{{c}^{\prime}(z)c(z)}{1-c^2(z)},
\end{equation}
where ${c}^{\prime}=\dot{c}/H$  and the prime represents
derivative with respect to $x=\ln a$. Combining Eqs.
(\ref{rhodot}) and (\ref{Hdot1})  with Eq. (\ref{ConserveDE}), one
can obtain the EoS parameter of GHDE as
\begin{equation}\label{wD1}
\omega_D=-\frac{2}{3}\frac{{c}^{\prime}(z)}{c(z)[1-c^2(z)]},
\end{equation}
Clearly, for constant $c$ parameter we have, ${c}^{\prime}(z)=0$
which leads to a wrong equation of state, namely that for dust
with $ \omega_D=0 $\cite{Hsu}. This implies that the HDE model
with $L=H^{-1}$ as IR cutoff cannot describe an accelerating
universe \cite{Hsu}. In contrast, taking the time varying $c^2$
term in the energy density of the HDE, it is quite possible to
reproduce the acceleration of the cosmic expansion in HDE model
with the  Hubble radius as IR cutoff.

Another important cosmological parameter for studying the
evolution of the universe is the deceleration parameter  which is
given by
\begin{equation}\label{q1}
q=1-\frac{\dot{H}}{H^2}.
\end{equation}
Substituting Eq. (\ref{Hdot1}) into (\ref{q1}) yields
\begin{equation}\label{q2}
q=\frac{1}{2}-\frac{{c}^{\prime}(z)c(z)}{1-c^2(z)}.
\end{equation}
Again for $c^{\prime}(z)=0$ the declaration parameter reduces to
$q={1}/{2} >0$ which implies a decelerated universe.

We see from Eqs. (\ref{wD1}) and (\ref{q2}) that the evolution of
these cosmological parameters depend on the functional form of $
c(z)$. In what follow, we consider four types of parametrization
for $ c(z) $ as given in Eqs. (\ref{G1})-(\ref{G4}).
\subsection{GHDE1: The CPL type}
We start with the CPL type for $c(z)$, namely
\begin{equation} \label{GHDE1}
c(z)=c_0+c_1\frac{z}{(1+z)}.
\end{equation}
When $ z\rightarrow \infty $ (in the early universe), we see that
$ c\rightarrow c_0+c_1 $ and as $ z\rightarrow 0 $ (at the present
time), $ c\rightarrow c_0 $. Thus holographic parameter $ c $
varies slowly from $ c_0+c_1 $ to $ c_0 $ from past to the
present.

Using the fact that $ a/a_0=(1+z)^{-1} $, where $z$ is the
redshift parameter and  prime denotes the derivative with respect
to $x=\ln a$, we arrive at $ {c}^{\prime}=-(1+z){dc}/{dz} $.
Taking derivatives of Eq. (\ref{GHDE1}), it follows that
\begin{equation} \label{dotGHDE1}
{c}^{\prime}=-(1+z)\frac{dc}{dz}=-\frac{c_1}{1+z}.
\end{equation}
Substituting Eqs. (\ref{GHDE1}) and (\ref{dotGHDE1}) into
(\ref{wD1}), the EoS parameter is obtained as
\begin{equation} \label{wGHDE1}
\omega_D
(z)=\frac{2}{3}\frac{c_1(1+z)^2}{[c_0(1+z)+c_1z](1+z)^2-[c_0(1+z)+c_1z]^3}.
\end{equation}
Combining Eqs. (\ref{GHDE1}) and (\ref{dotGHDE1}) with Eq.
(\ref{q2}), we can obtain the deceleration parameter as
\begin{equation} \label{qGHDE1}
q(z)=\frac{1}{2}+\frac{c_0c_1(1+z)+c_1^2z}{(1+z)^2-[c_0(1+z)+c_1z]^2}.
\end{equation}
The behavior of $ \omega_D (z) $ and $ q(z) $ are plotted for
different values of model parameters $ c_0 $ and $ c_1 $ in
Fig.\ref{w1,q1-H}. From these figures we see that our Universe has
a transition from deceleration to the acceleration phase around $
z\approx 0.6 $ which is consistent with observations
\cite{Daly,Kom1,Kom2,Planck1,Planck2}.
\begin{figure}[htp]
\begin{center}
\includegraphics[width=8cm]{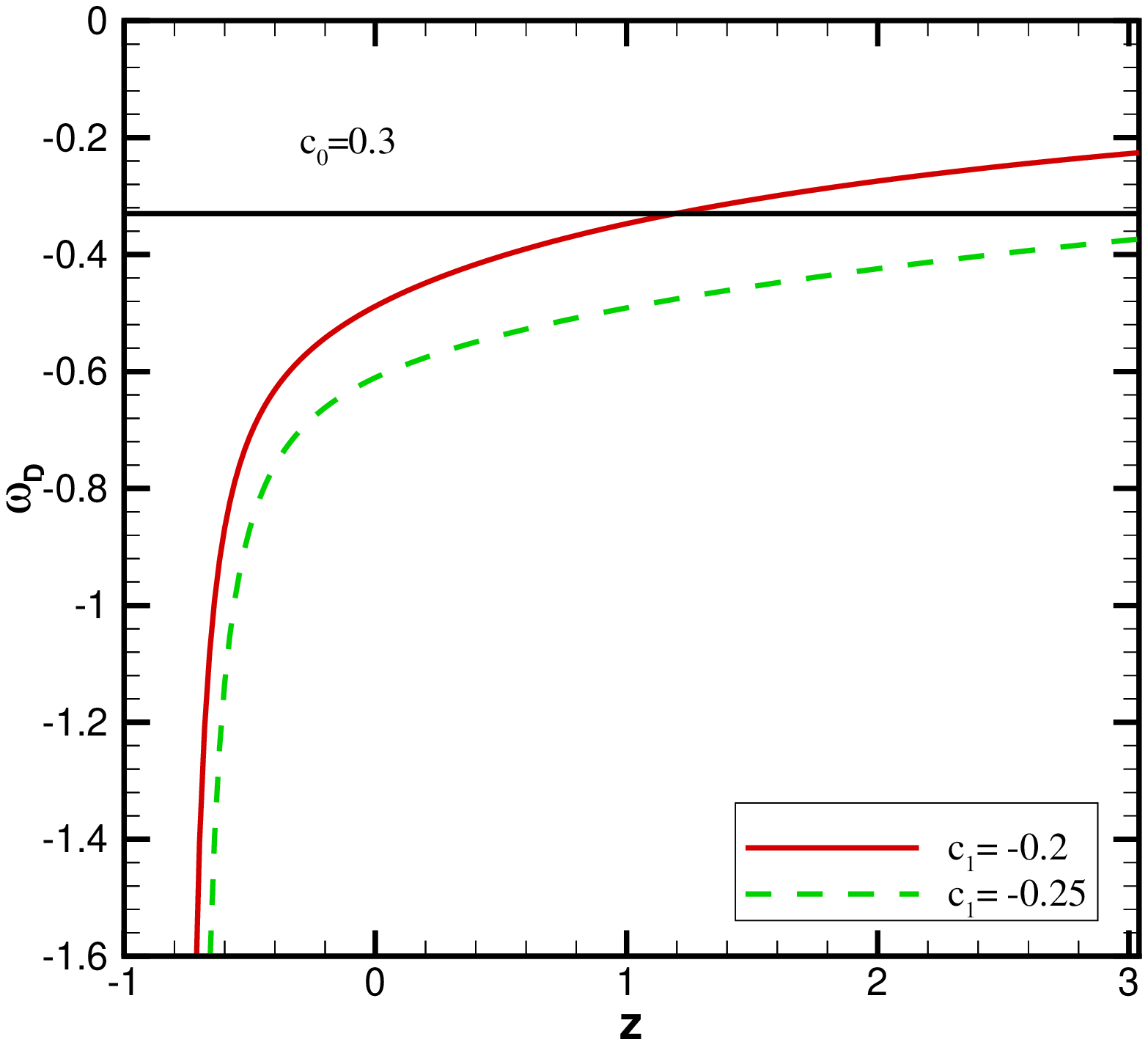}
\includegraphics[width=8cm]{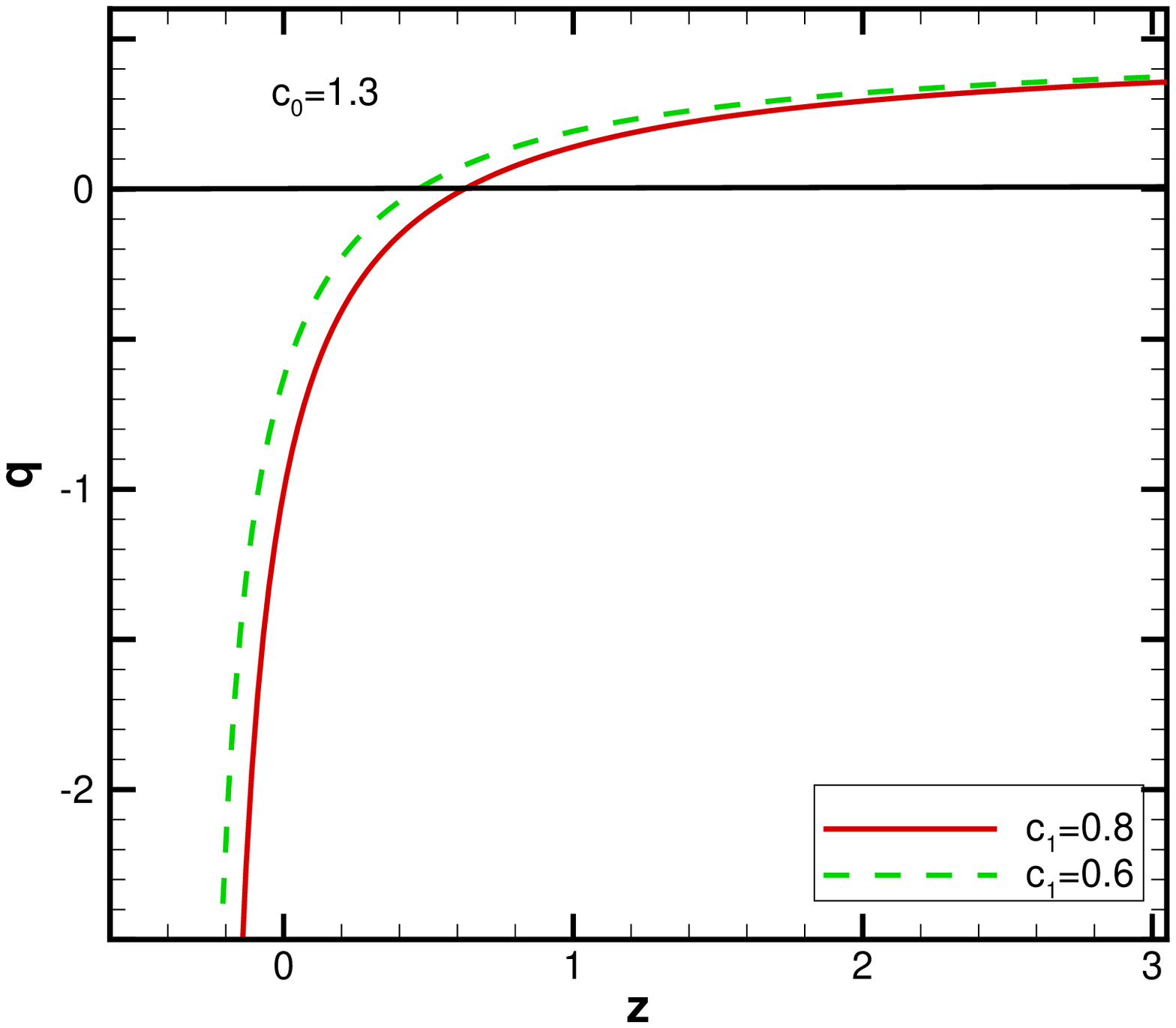}
\caption{The evolution of EoS parameter $ \omega_D $ and the
deceleration parameter $ q $
        versus redshift $ z $ for  GHDE1 model with $ L=H^{-1}.$ }\label{w1,q1-H}
\end{center}
\end{figure}
\subsection{GHDE2: The JBP model}
The second parametrization of  $c(z)$ is the JBP parametrization
which is written as
\begin{equation} \label{GHDE2}
c(z)=c_0+c_1\frac{z}{(1+z)^2}.
\end{equation}
From Eq. (\ref{GHDE2}) we see that at the late time where $
z\rightarrow 0 $, we have $ c(z)\rightarrow c_0 $,  and as $
z\rightarrow 1 $, we have $ c (z)\rightarrow c_0+{c_1}/{4} $.
Besides, in the early universe where $ z\rightarrow \infty $,  we
have $ c (z)\rightarrow c_0 $.

Taking derivative of Eq. (\ref{GHDE2}) we find
\begin{equation} \label{dotGHDE2}
{c}^{\prime}(z)=-c_1\frac{1-z}{(1+z)^2}.
\end{equation}
Substituting Eqs.  (\ref{GHDE2})  and (\ref{dotGHDE2}) into Eq.
(\ref{wD1}),  we obtain the EoS parameter for GHDE2 model as
\begin{equation} \label{wGHDE2}
\omega_D=\frac{2}{3}\frac{c_1(1-z)(1+z)^4}{[c_0(1+z)^2+c_1z][(1+z)^4-(c_0(1+z)^2+c_1z)^2]}.
\end{equation}
Combining Eqs. (\ref{GHDE2}) and (\ref{dotGHDE2}) with Eq.
(\ref{q2}) yields
\begin{equation} \label{qGHDE2}
q=\frac{1}{2}+\frac{c_1(1-z)[c_0(1+z)^2+c_1z]
}{(1+z)^4-[c_0(1+z)^2+c_1z]^2}.
\end{equation}
The behavior of $ \omega_D (z) $ and $ q(z) $ are plotted for
different values of model parameters $ c_0 $ and $ c_1 $ in Fig.
\ref{w2,q2-H}. Again, the universe has a transition from
deceleration to the acceleration phase around $ z\approx 0.6 $ and
at the late time where $z\rightarrow0$, the EoS parameter can
cross the phantom line $w_D=-1$.
\begin{figure}[htp]
\begin{center}
\includegraphics[width=8cm]{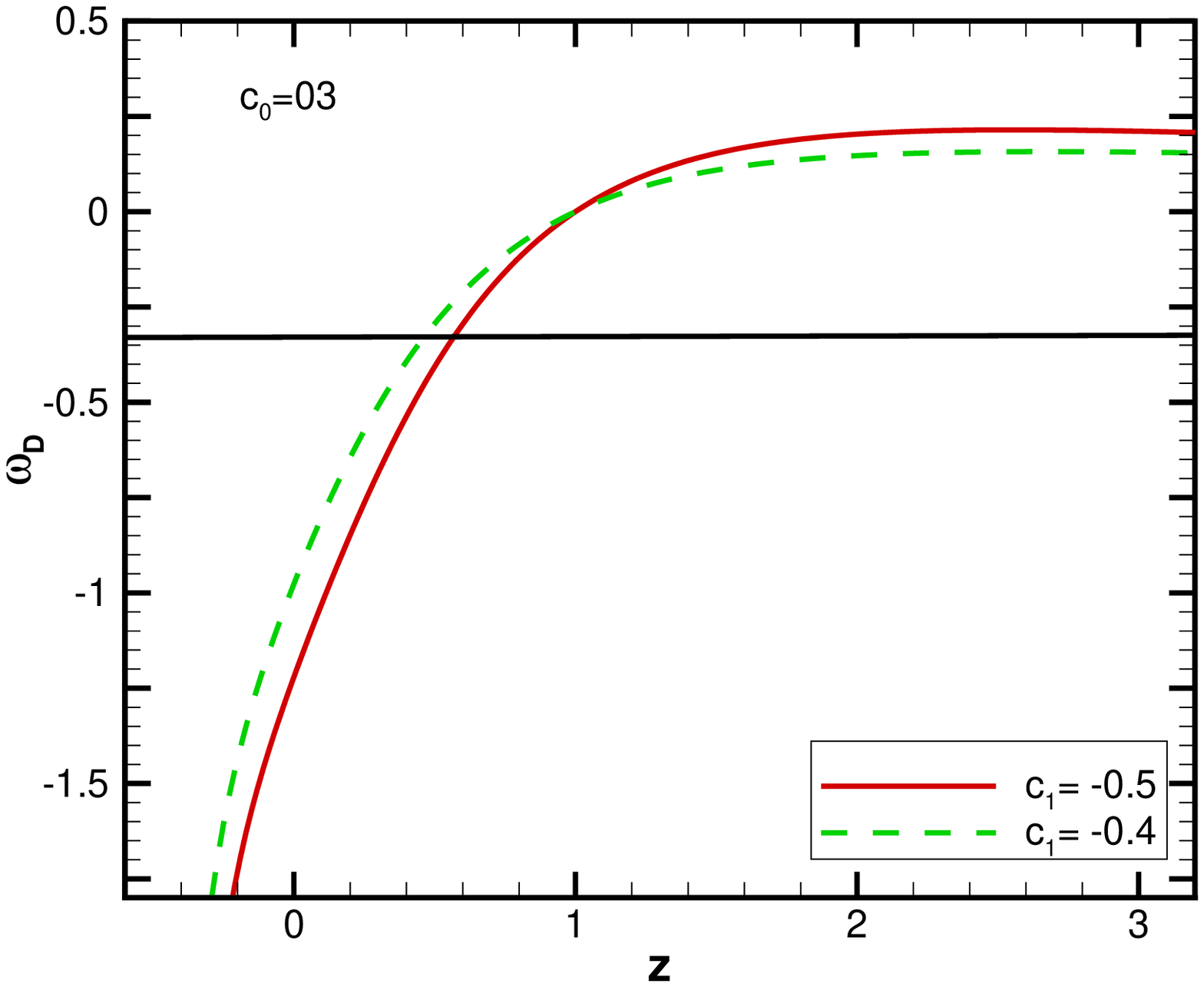}
\includegraphics[width=8cm]{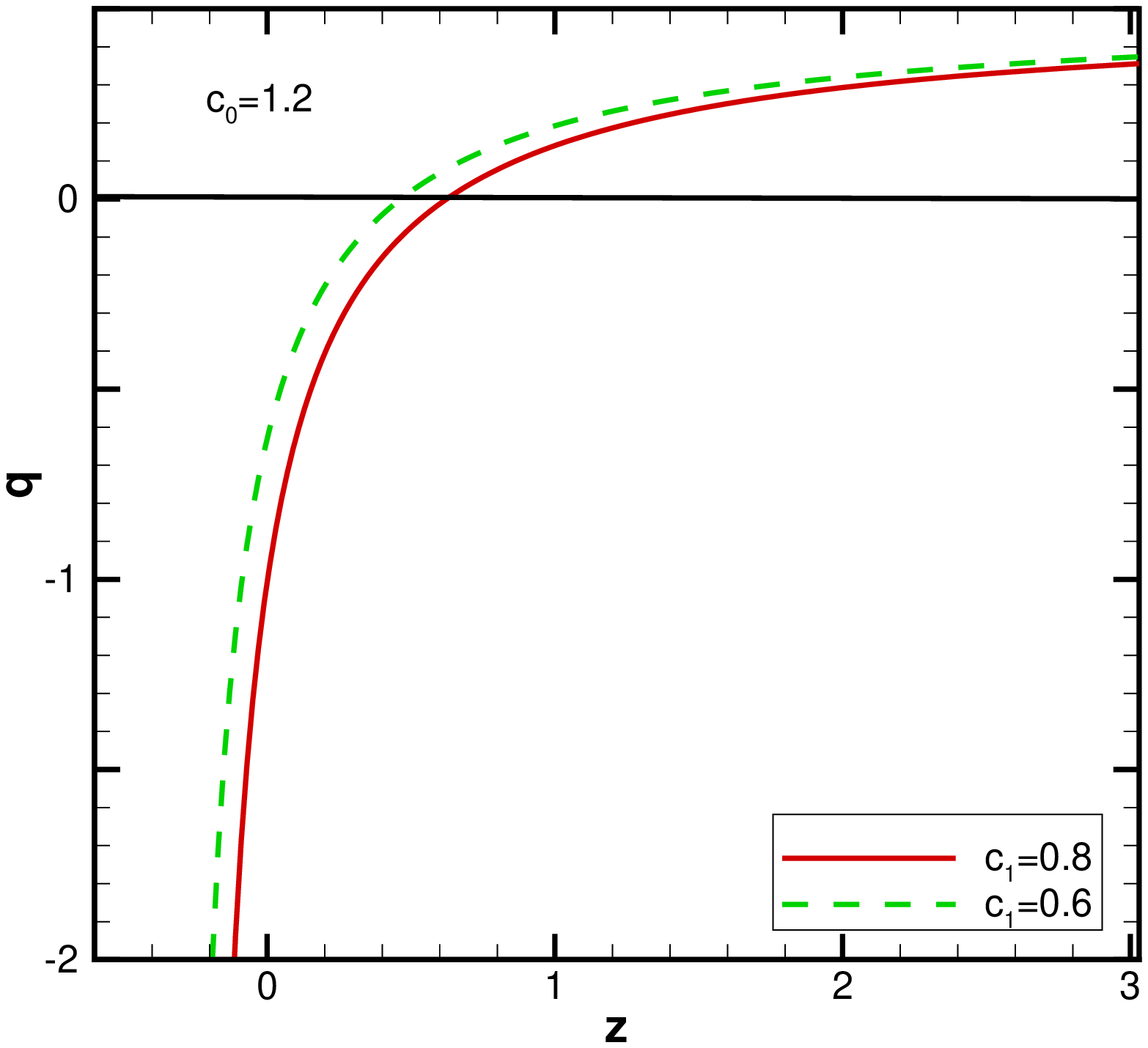}
\caption{The evolution of EoS parameter $ \omega_D $ and the
deceleration parameter $ q $
        versus redshift $ z $ for  GHDE2 model with $ L=H^{-1}.$ }\label{w2,q2-H}
\end{center}
\end{figure}
\subsection{GHDE3: The Wetterich type }
The third parametrization is Wetterich-type which assumes the
following form of $c(z)$:
\begin{equation} \label{GHDE3}
c(z)=\frac{c_0}{1+c_1\ln(1+z)}.
\end{equation}
In this model, at the late time where $ z\rightarrow 0 $, we have
$ c(z)\rightarrow c_0 $, while at the early universe where $
z\rightarrow \infty $, we get $ c(z) \rightarrow 0 $ and thus
$\rho_D\rightarrow0$. This implies that at the early universe the
HDE did not have significant contribution in the evolution of the
universe. It follows directly that,
\begin{equation} \label{dotGHDE3}
{c}^{\prime}(z)=-(1+z)\frac{d
c(z)}{dz}=\frac{c_0c_1}{[1+c_1\ln(1+z)]^2}.
\end{equation}
Inserting Eqs. (\ref{GHDE3}) and (\ref{dotGHDE3}) into
(\ref{wD1}), one gets
\begin{equation} \label{wGHDE3}
\omega_D(z)=-\frac{2}{3}\frac{c_1[1+c_1\ln(1+z)]}{[1+c_1\ln(1+z)]^2-c_0^2}.
\end{equation}
Combining Eqs. (\ref{q2}), (\ref{GHDE3}) and (\ref{dotGHDE3}), we
find
\begin{equation} \label{qGHDE3}
q(z)=\frac{1}{2}+\frac{c_0^2c_1}{[1+c_1\ln(1+z)]\Big(c_0^2-[1+c_1\ln(1+z)]^2\Big)}.
\end{equation}
We have plotted the evolutions of $ \omega_D (z) $ and $ q(z) $ in
terms of the redshift parameter $z$ in Fig. \ref{w3,q3-H}. From
these figures it is obvious that the present acceleration can be
addressed in this model and the transition from deceleration to
the acceleration phase occurs for $ 0.3<z< 0.7 $.
\begin{figure}[htp]
\begin{center}
\includegraphics[width=8cm]{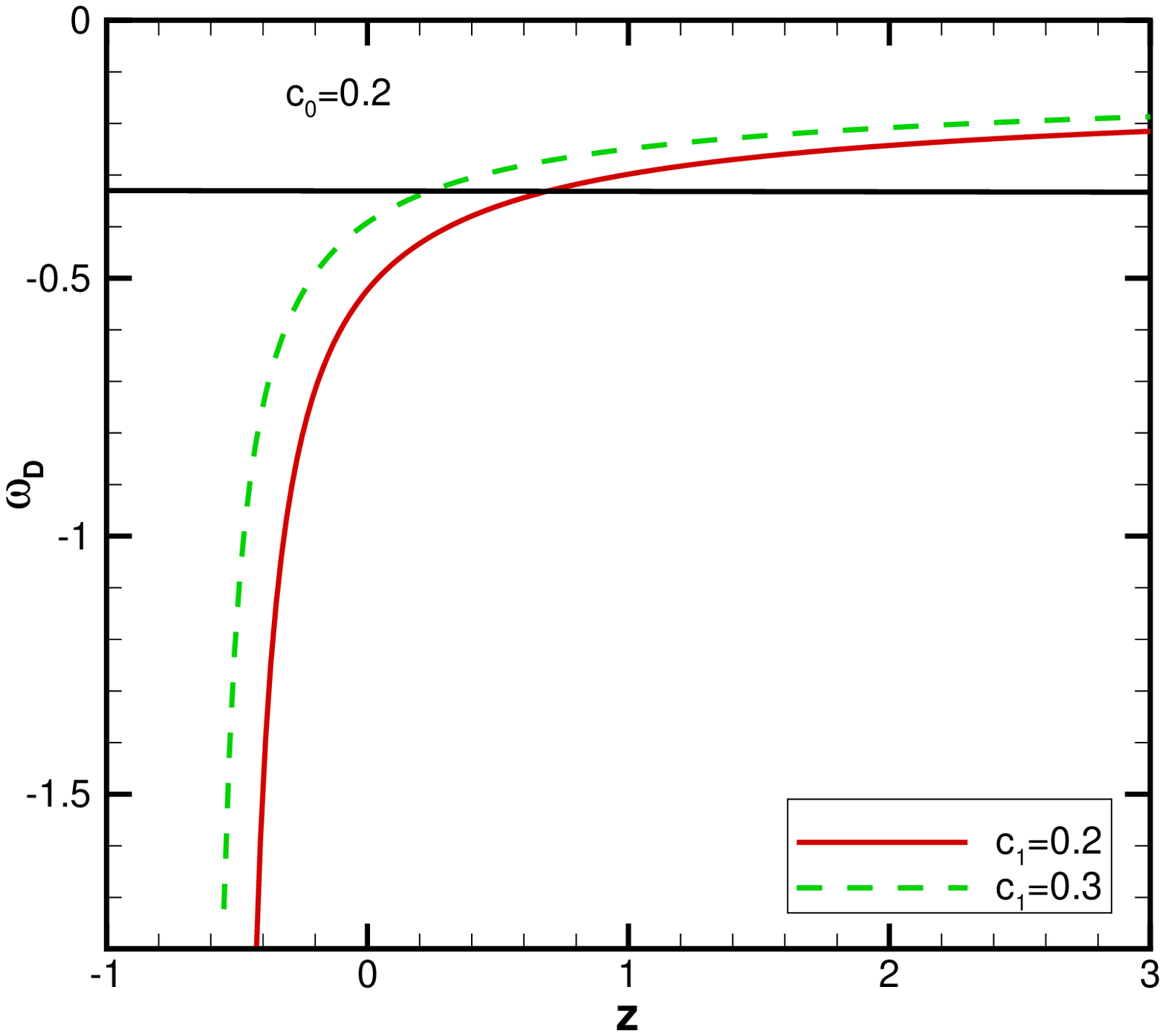}
\includegraphics[width=8cm]{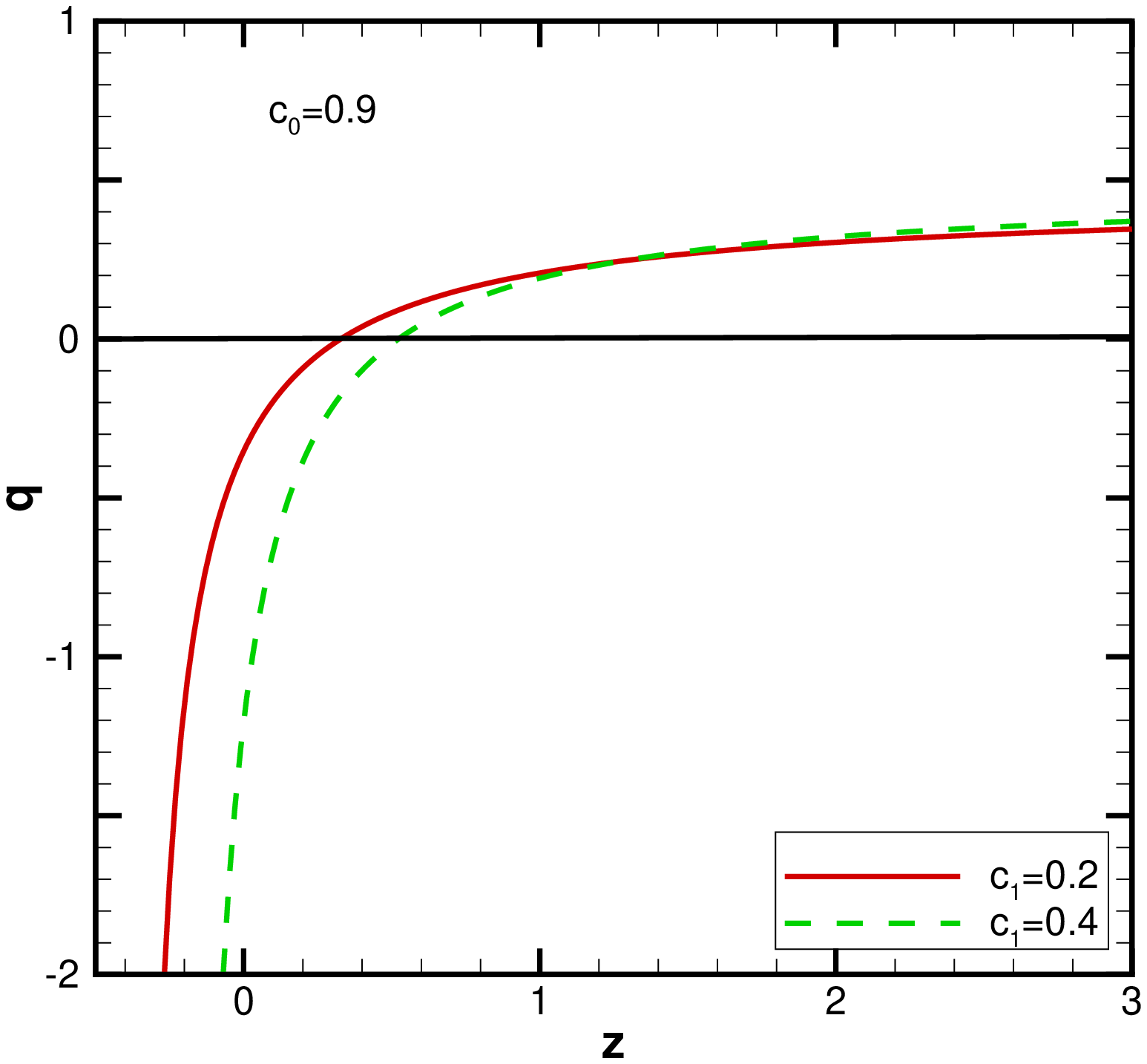}
\caption{The evolution of $ \omega_D $ and $ q $
        versus redshift $ z $ for  GHDE3 with $ L=H^{-1}.$ }\label{w3,q3-H}
\end{center}
\end{figure}
\subsection{GHDE4: The Ma-Zhang type}
The last choice for the parametrization of $c(z)$ was proposed in
\cite{Ma} and can be written as
\begin{equation}\label{GHDE4}
c(z)=c_0+c_1\left(\frac{\ln(2+z)}{1+z}-\ln 2\right).
\end{equation}
At the present time where $ z\rightarrow 0 $ we have $
c(z)\rightarrow c_0 $, and at the early time where $z\rightarrow
\infty $, one gets $ c(z) \rightarrow c_0-c_1\ln 2$. It is worth
noting that for the previous choice of $c(z)$, we could not
investigate the future behavior of $c(z)$, because it diverges at
the future time  where $ z\rightarrow -1 $. However, in case of
the Ma-Zhang parametrization we have  $ c(z)\rightarrow
c_0+c_1(1-\ln 2) $ as $ z\rightarrow -1 $. From  Eq. (\ref{GHDE4})
it is easy to show that
\begin{equation}\label{dotGHDE4}
{c}^{\prime}(z)=-(1+z)\frac{d
c(z)}{dz}=c_1\frac{(2+z)\ln(2+z)-(1+z)}{(1+z)(2+z)}.
\end{equation}
Combining Eqs. (\ref{GHDE4}) and (\ref{dotGHDE4}) with (\ref{wD1})
and (\ref{q2}), we arrive at
\begin{equation}\label{wGHDE4}
\omega_D(z)=\frac{2}{3}\frac{c_1(1+z)^2[(1+z)-(2+z)\ln(2+z)]}
{(2+z)\Big(c_0(1+z)+c_1\ln(2+z)-c_1(1+z)\ln
2\Big)\Big[(1+z)^2-[c_0(1+z)+c_1\ln(2+z)-c_1(1+z)\ln 2]^2\Big]},
\end{equation}
\begin{equation} \label{qGHDE4}
q(z)=\frac{1}{2}+\frac{c_1[c_0(1+z)+c_1\ln(2+z)-c_1(1+z)\ln
2][1+z-(2+z)\ln(2+z)]}
{(2+z)\Big((1+z)^2-[c_0(1+z)+c_1\ln(2+z)-c_1(1+z)\ln 2]^2\Big)}.
\end{equation}
In order to have an insight on the behaviour of these functions,
we plot them in terms of $z$  in Fig. \ref{w4,q4-H}. From these
figures, we see that the behaviour is similar to the previous
parameterizations.

\begin{figure}[htp]
\begin{center}
\includegraphics[width=8cm]{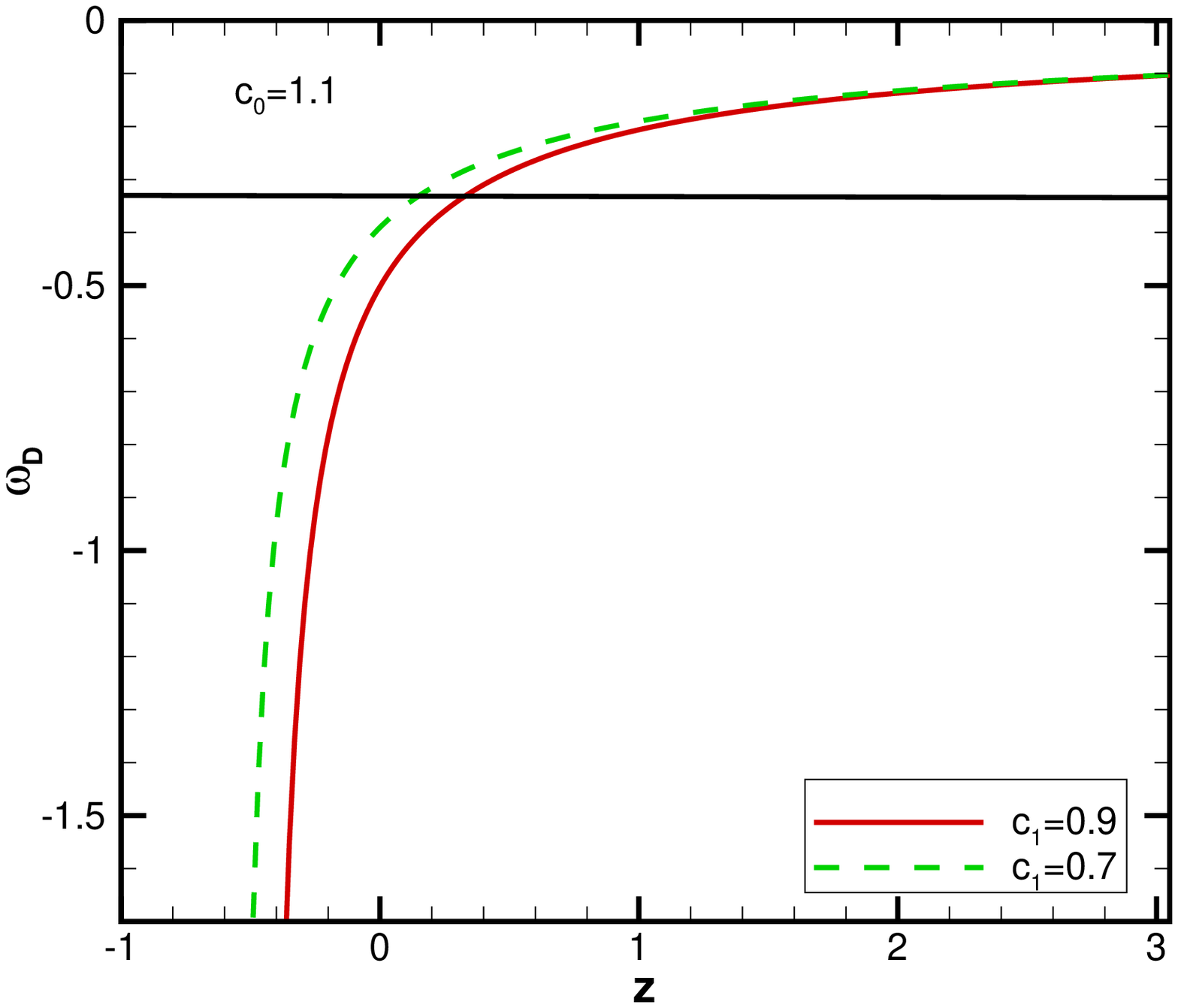}
\includegraphics[width=8cm]{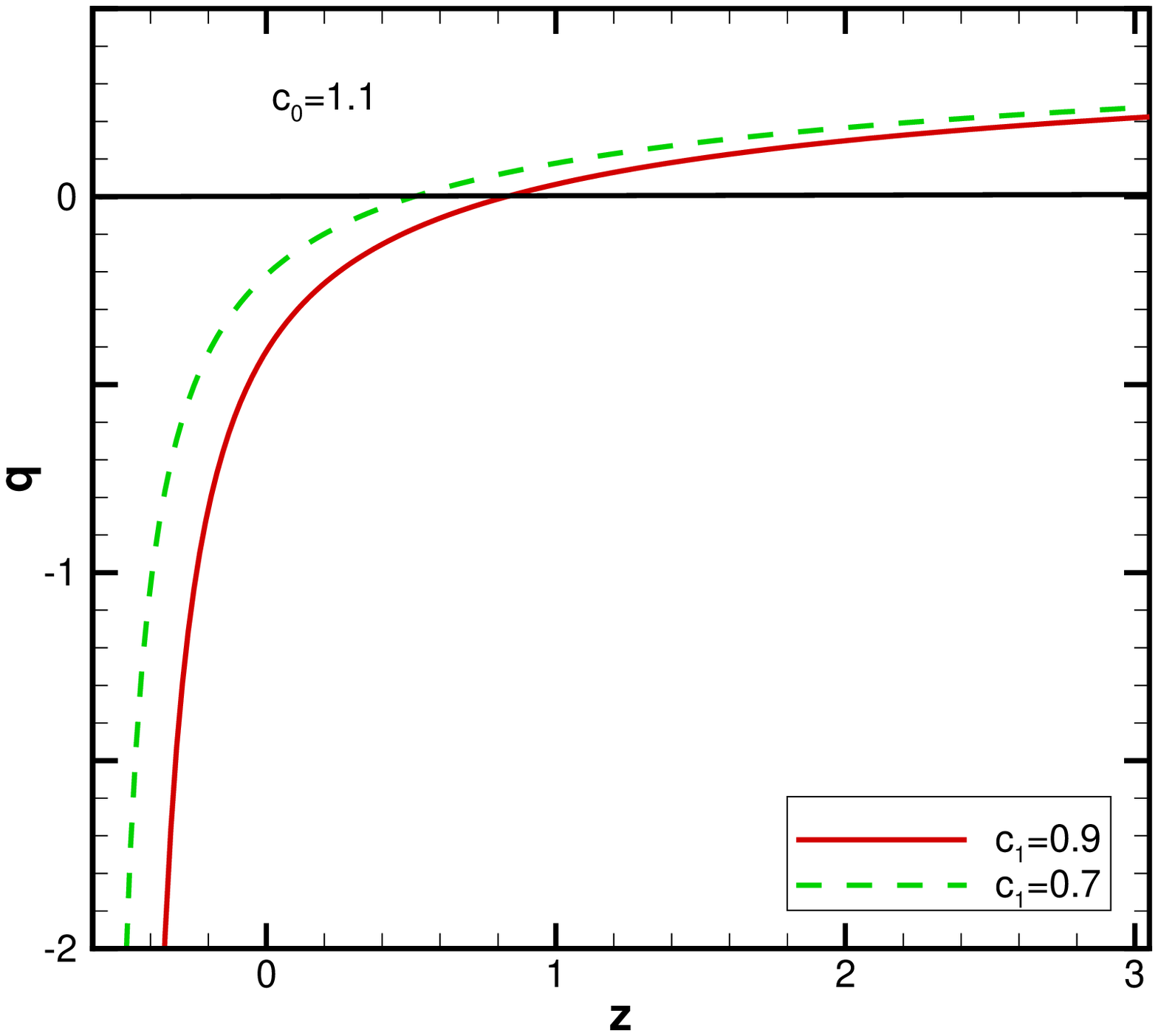}
\caption{The evolution of EoS parameter $ \omega_D $ and the
deceleration parameter $ q $
        versus redshift $ z $ for  GHDE4 with $ L=H^{-1}.$ }\label{w4,q4-H}
\end{center}
\end{figure}
\section{GHDE in flat universe with GO cutoff}
In this section we consider the GO cutoff as system's IR cutoff,
namely $ L=(\alpha H^2+\beta\dot{H})^{{-1}/{2}} $ which first
proposed in \cite{Granda}. With this IR cutoff, the energy density
(\ref{HDE}) is written
\begin{equation}\label{GOHDE}
\rho_D=3M_p^2c^2(z)\left(\alpha H^2+\beta\dot{H}\right),
\end{equation}
where $ \alpha $ and $ \beta $ are constants that should be
constrained by observational data. Using the definition of density
parameter (\ref{Friedeq2}) one can obtain
\begin{equation}\label{Friedeq4}
\Omega_D=c^2(z)\left(\alpha +\beta \frac{\dot{H}}{H^2}\right).
\end{equation}
Taking the time derivative of Eq. (\ref{Friedeq4}), we get
\begin{equation}\label{Omegadot}
\dot{\Omega}_D=2\Omega_D\left(\frac{\dot{c}(z)}{c(z)}-\frac{\dot{H}}{H}\right)+
\frac{c^2(z)}{H^2}\left(2\alpha \dot{H}H +\beta \ddot{H}\right).
\end{equation}
Now, if we take the time derivative of both sides of Friedmann
equation (\ref{Friedeq1}) and after using Eq. (\ref{GOHDE}), we
arrive at
\begin{equation}\label{ddH}
c^2H^{-3}(2\alpha\dot{H}H +\beta
\ddot{H})=2\frac{\dot{H}}{H^2}+3(1-\Omega_D)-2\Omega_D\frac{\dot{c}(z)}{Hc(z)}.
\end{equation}
Combining Eqs. (\ref{Omegadot}) and (\ref{ddH}) the equation of
motion for the dimensionless GHDE density  can be written as
\begin{equation}\label{Omegadot2}
\dot{\Omega}_D=(1-\Omega_D)\left(\dfrac{2\dot{H}}{H}+3H\right).
\end{equation}
By help of Eq. (\ref{Friedeq4}) and using the fact that $
\dot{\Omega}_D=H{\Omega}^{\prime}_D $,  the evolution of
dimensionless GHDE density may be rewritten as
\begin{equation}\label{Omegaprime}
{\Omega}^{\prime}_D(z)=(1-\Omega_D)\left(\frac{2\Omega_D}{\beta
c^2(z)}-\frac{2\alpha}{\beta}+3\right),
\end{equation}
where the prime denotes derivative with respect to $ x=\ln{a} $.
Taking the time derivative of both sides of Eq. (\ref{Friedeq1})
and using Eqs. (\ref{ConserveDE}) and (\ref{Friedeq4}) we can
obtain the EoS parameter of GHDE model as follows
\begin{equation}\label{wD2}
\omega_D(z)=\frac{2\alpha}{3\beta \Omega_D}-\frac{2}{3\beta
c^2(z)}-\frac{1}{\Omega_D}.
\end{equation}
Substituting Eq. (\ref{Friedeq4}) into (\ref{q1}), we get
\begin{equation}\label{q3}
q(z)=-1-\frac{\Omega_D}{\beta c^2(z)}+\frac{\alpha}{\beta}.
\end{equation}
Following the previous section, we shall consider four types of
parametrization of $c(z) $ listed in Eqs. (\ref{G1})-(\ref{G4}),
respectively.
\begin{figure}[htp]
\begin{center}
\includegraphics[width=8cm]{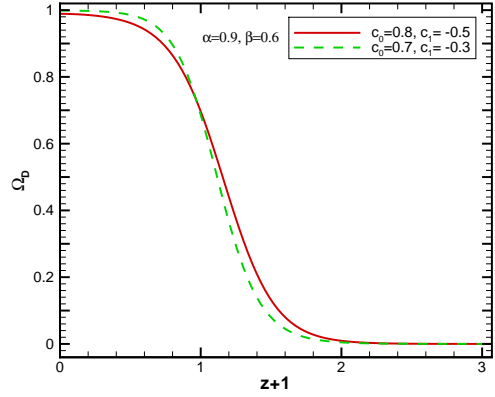}
\caption{The evolution of the dimensionless density parameter $
\Omega_D $
        versus redshift $ z $ for  GHDE1 model with GO cutoff. }\label{Omega1}
\end{center}
\end{figure}
\subsection{GHDE1: The CPL type}
Substituting Eq. (\ref{G1}) in (\ref{Omegaprime}), the equation of
the evolutionary of dimensionless GHDE1 density is obtained as
\begin{equation}\label{Omegaprime1}
{\Omega}^{\prime}_D=(1-\Omega_D)\left(\frac{2\Omega_D(1+z)^2}{\beta[c_0(1+z)+c_1z]^2}-\frac{2\alpha}{\beta}+3\right)
\end{equation}
The evolution of the dimensionless GHDE density parameter $
\Omega_D $ as a function of $1+z = a^{-1}$ is shown in Fig.
\ref{Omega1}. From this figure we see that at the early universe
where $ z\rightarrow \infty $ we have  $ \Omega_D\rightarrow 0$,
while at the late time where $ z\rightarrow -1 $, the DE dominated, namely $ \Omega_D\rightarrow 1 $.\\
Using Eqs. (\ref{G1}), (\ref{wD2}) and (\ref{q3}) we can obtain
the EoS and the deceleration parameters as
\begin{eqnarray}\label{wGO1}
\omega_D(z)&=&\frac{2\alpha}{3\beta
\Omega_D}-\frac{2(1+z)^2}{3\beta
[c_0(1+z)+c_1z]^2}-\frac{1}{\Omega_D}, \\
q(z)&=&-1-\frac{\Omega_D(1+z)^2}{\beta(c_0(1+z)+c_1z)^2}+\frac{\alpha}{\beta}.\label{qGO1}
\end{eqnarray}
The evolution of EoS parameter $ \omega_D (z) $ and deceleration
parameter $ q(z) $ are shown numerically in Fig. \ref{w1,q1-GO}.
where we have fixed $ c_0=2.7, c_1=-0.7 $ for different values of
$ \alpha$ and $\beta $. From these figures we clearly see that we
have a transition from a decelerated to an accelerated universe
around $z\approx 0.6$  which is compatible with observations
\cite{Planck1,Planck2}. However, not only at the present time
($z=0$), but also at the far future where $1+z\rightarrow0$, the
EoS parameter, $\omega_D$, cannot cross the phantom line and we
have always $\omega_D>-1$.
\begin{figure}[htp]
\begin{center}
\includegraphics[width=8cm]{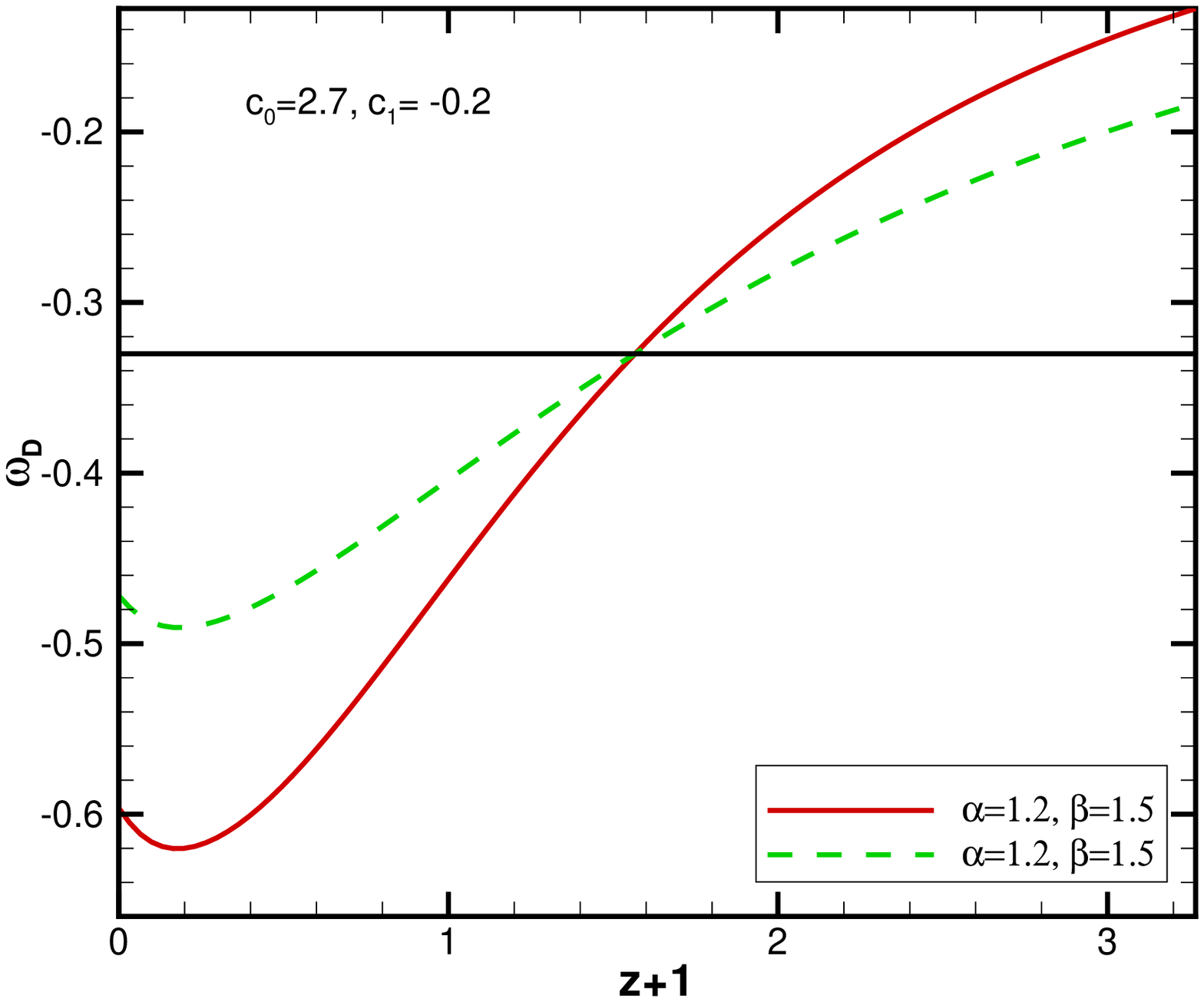}
\includegraphics[width=8cm]{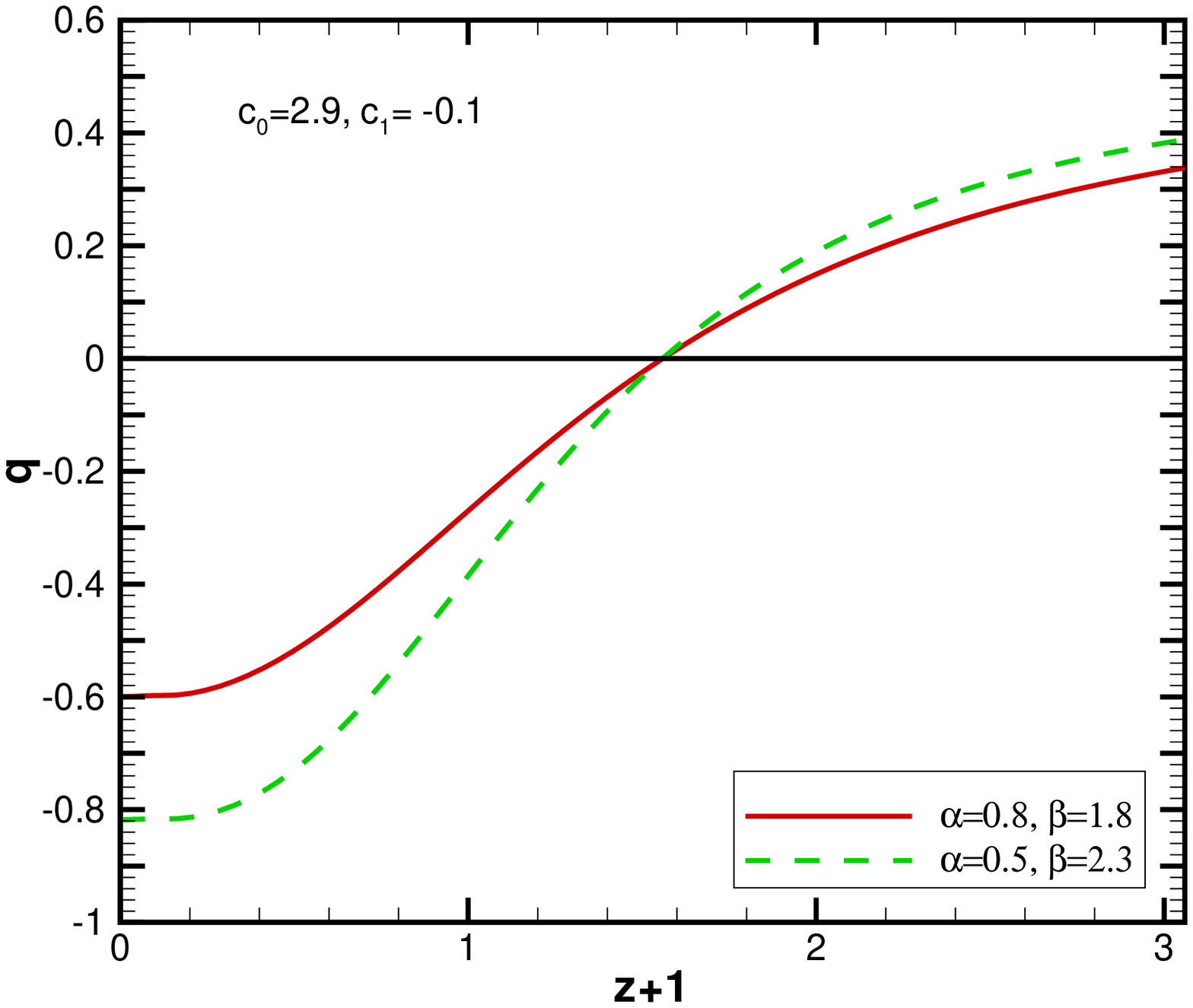}
\caption{The evolution of equation of state parameter $ \omega_D $
and deceleration parameter $ q $
        versus $1+z $ for  GHDE1 model with GO cutoff. }\label{w1,q1-GO}
\end{center}
\end{figure}
\subsection{GHDE2: The JBP model}
\begin{figure}[htp]
\begin{center}
\includegraphics[width=8cm]{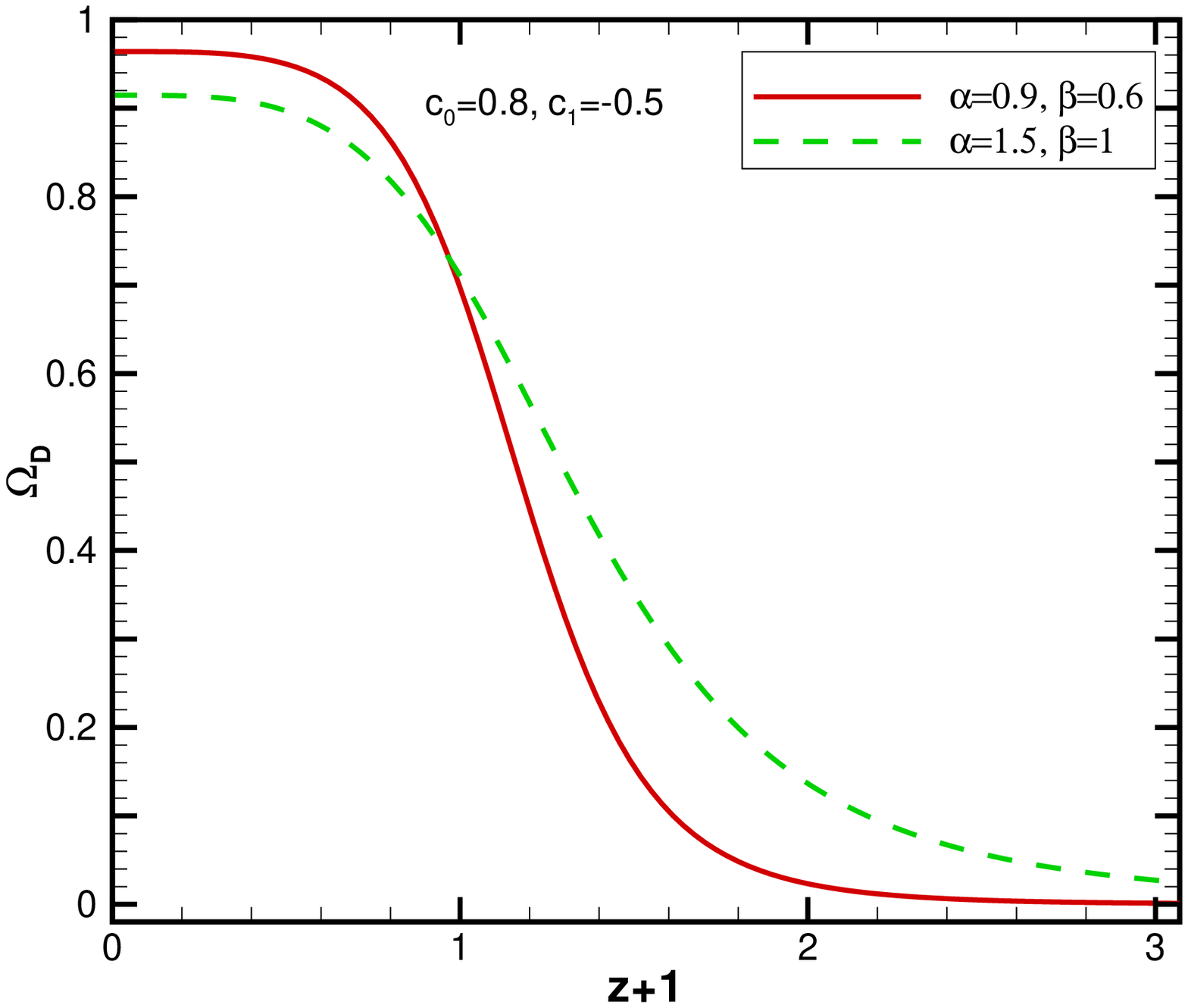}
\caption{The evolution of the dimensionless density parameter $
\Omega_D $
        versus  $ 1+z $ for  GHDE2 model with GO cutoff.}\label{Omega2}
\end{center}
\end{figure}

Combining Eqs. (\ref{G2}) and (\ref{Omegaprime}) one can derive
the evolution of dimensionless GHDE2 density as
\begin{equation}\label{Omegaprime2}
{\Omega}^{\prime}_D=(1-\Omega_D)\left(\frac{2\Omega_D(1+z)^4}{\beta(c_0(1+z)^2+c_1z)^2}-\frac{2\alpha}{\beta}+3\right).
\end{equation}
The evolution of the dimensionless GHDE density parameter $
\Omega_D $ as a function of redshift $ z $ is shown in Fig.
\ref{Omega2}.
Using Eqs. (\ref{G2}), (\ref{wD2}) and (\ref{q3}) we can obtain
the equation of state and deceleration parameters as
\begin{eqnarray}\label{wGO2}
\omega_D(z)&=&\frac{2\alpha}{3\beta
\Omega_D}-\frac{2(1+z)^4}{3\beta
[c_0(1+z)^2+c_1z]^2}-\frac{1}{\Omega_D}, \\
\label{qGO2}
q(z)&=&-1-\frac{\Omega_D(1+z)^4}{\beta[c_0(1+z)^2+c_1z]^2}+\frac{\alpha}{\beta}.
\end{eqnarray}
The behavior of the EoS parameter $\omega_D$ and the deceleration
parameter $ q $ are plotted in Fig. \ref{w2,q2-GO}. Again,  our
universe has a phase transition during its history from a
deceleration to an accelerated phase and $\omega_D$ cannot cross
the phantom line even in the far future where $1+z\rightarrow0$.
\begin{figure}[htp]
\begin{center}
\includegraphics[width=8cm]{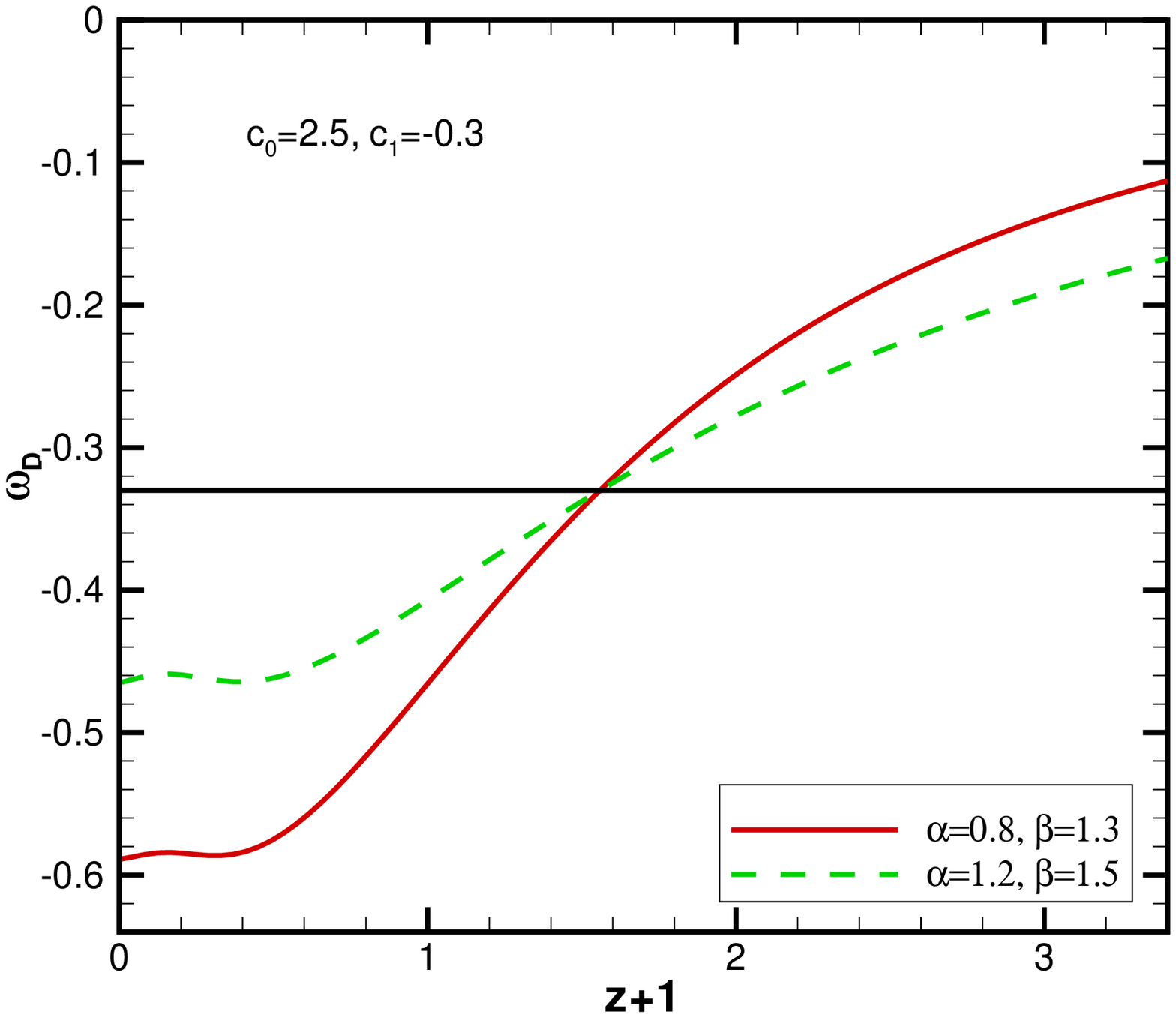}
\includegraphics[width=8cm]{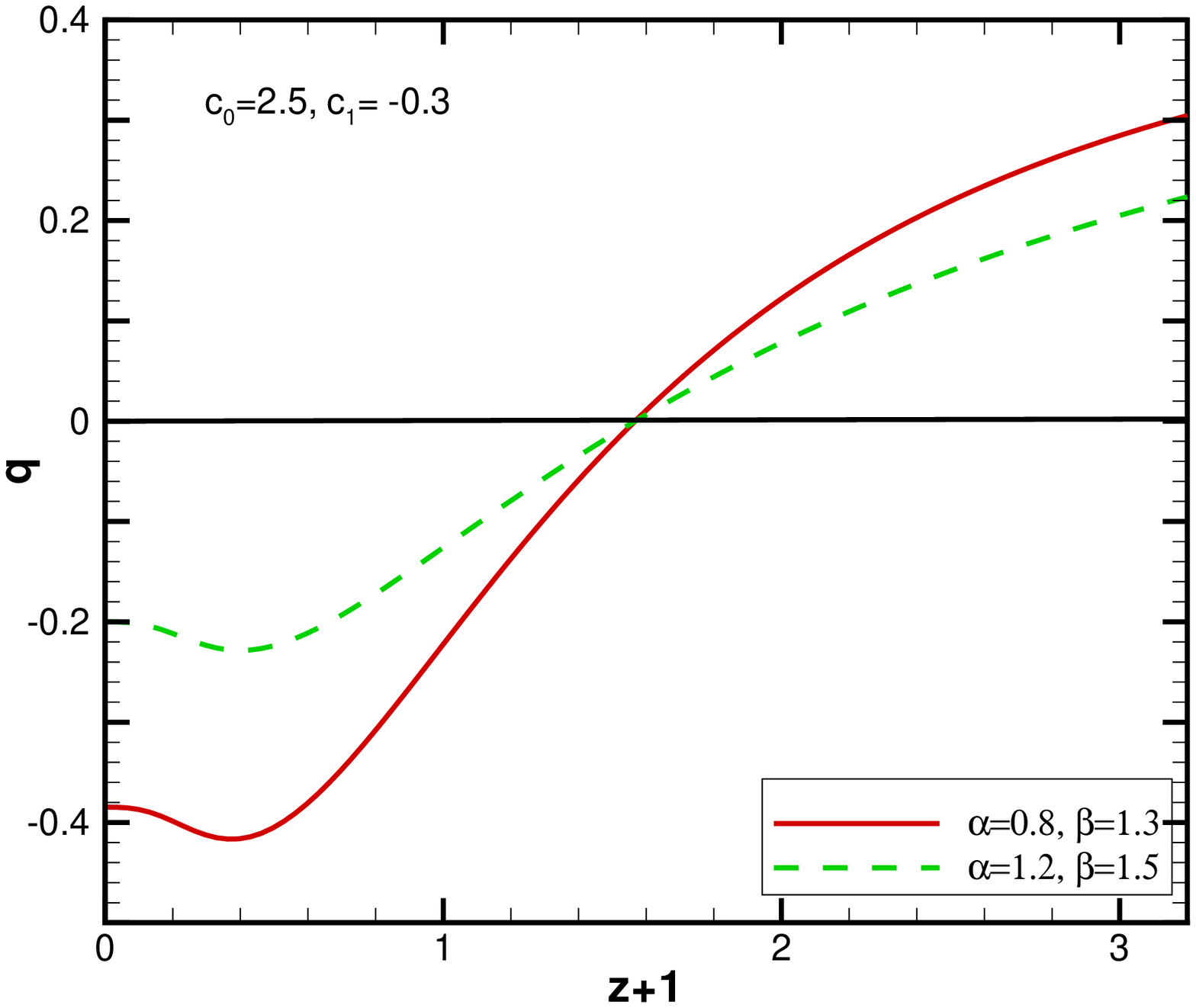}
\caption{The evolution of equation of state parameter $ \omega_D $
and deceleration parameter $ q $
        versus  $1+z $ for  GHDE2 model with GO cutoff. }\label{w2,q2-GO}
\end{center}
\end{figure}
\subsection{GHDE3: The Wetterich type}
\begin{figure}[htp]
\begin{center}
\includegraphics[width=8cm]{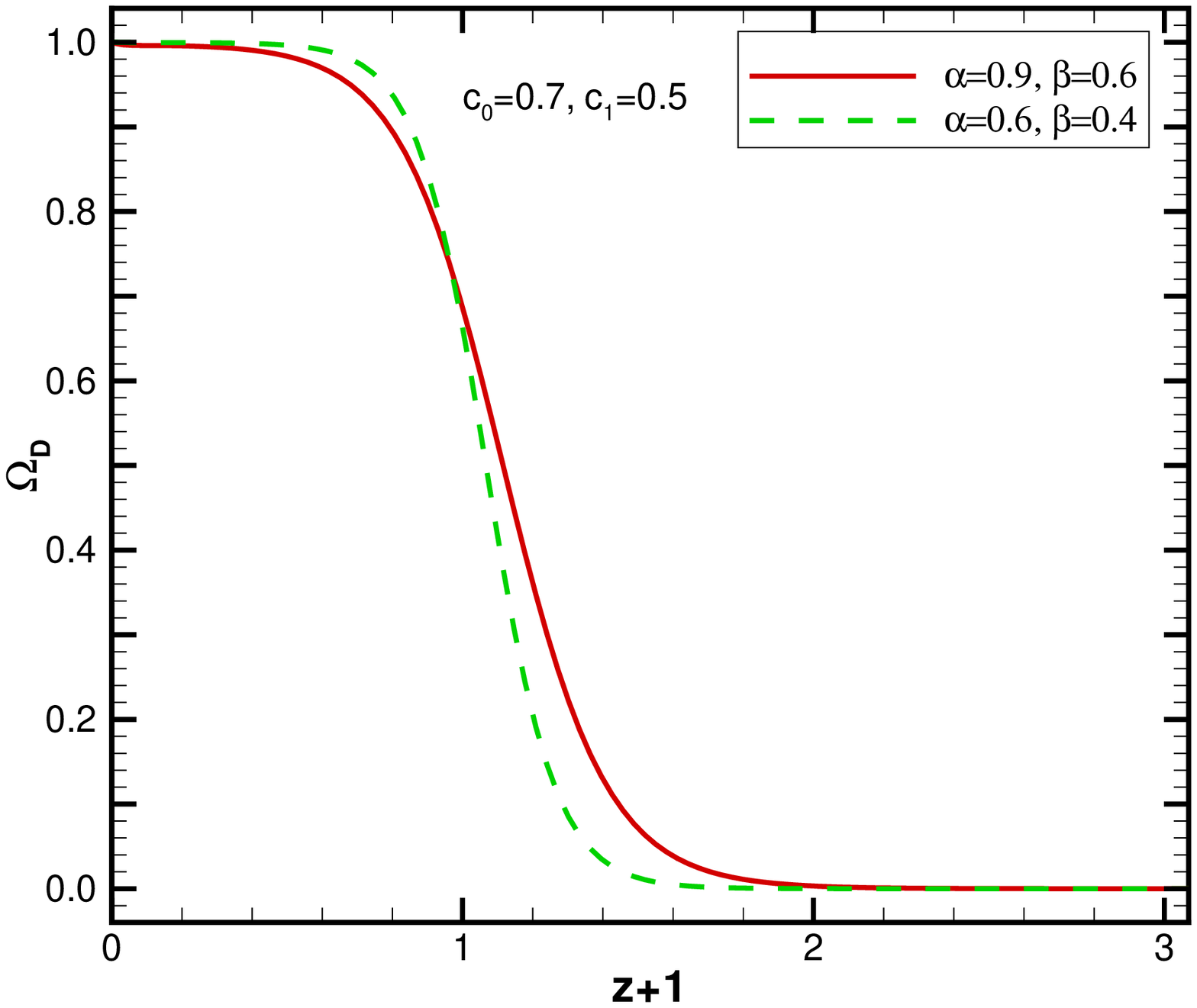}
\caption{The evolution of the dimensionless density parameter $
\Omega_D $
        versus redshift $1+z $ for GHDE3 model with GO cutoff.}\label{Omega3}
\end{center}
\end{figure}

Using Eqs. (\ref{G3}) and (\ref{Omegaprime}) we obtain the
evolution of dimensionless GHDE3 density as
\begin{equation}\label{Omegaprime3}
{\Omega}^{\prime}_D=(1-\Omega_D)\left(\frac{2\Omega_D(1+c_1\ln(1+z))^2}{\beta
c_0^2}-\frac{2\alpha}{\beta}+3\right).
\end{equation}
The evolution of the dimensionless GHDE density parameter $
\Omega_D $ as a function of redshift $ z $ is shown in Fig.
(\ref{Omega3}).
Inserting Eq. (\ref{G3}) in Eqs. (\ref{wD2}) and (\ref{q3}) we can
obtain the EoS and the deceleration parameters as follows
\begin{eqnarray}\label{wGO3}
\omega_D(z)&=&\frac{2\alpha}{3\beta
\Omega_D}-\frac{2[1+c_1\ln(1+z)]^2}{3\beta
c_0^2}-\frac{1}{\Omega_D},\\
q(z)&=&-1-\frac{\Omega_D[1+c_1\ln(1+z)]^2}{\beta
c_0^2}+\frac{\alpha}{\beta}. \label{qGO3}
\end{eqnarray}
The behavior of the EoS parameter $\omega_D$ and deceleration
parameter $ q $ are plotted in Fig. \ref{w3,q3-GO}. From these
figures we see that, in contrast to the two previous
parametrization of $c(z)$, which $\omega_D$ cannot cross the
phantom line and $\omega_D>-1$ as $1+z\rightarrow0$, here for the
Wetterich type of parametrization, the EoS parameter can cross the
phantom line ($\omega_D<-1$)  as $1+z\rightarrow0$.
\begin{figure}[htp]
\begin{center}
\includegraphics[width=8cm]{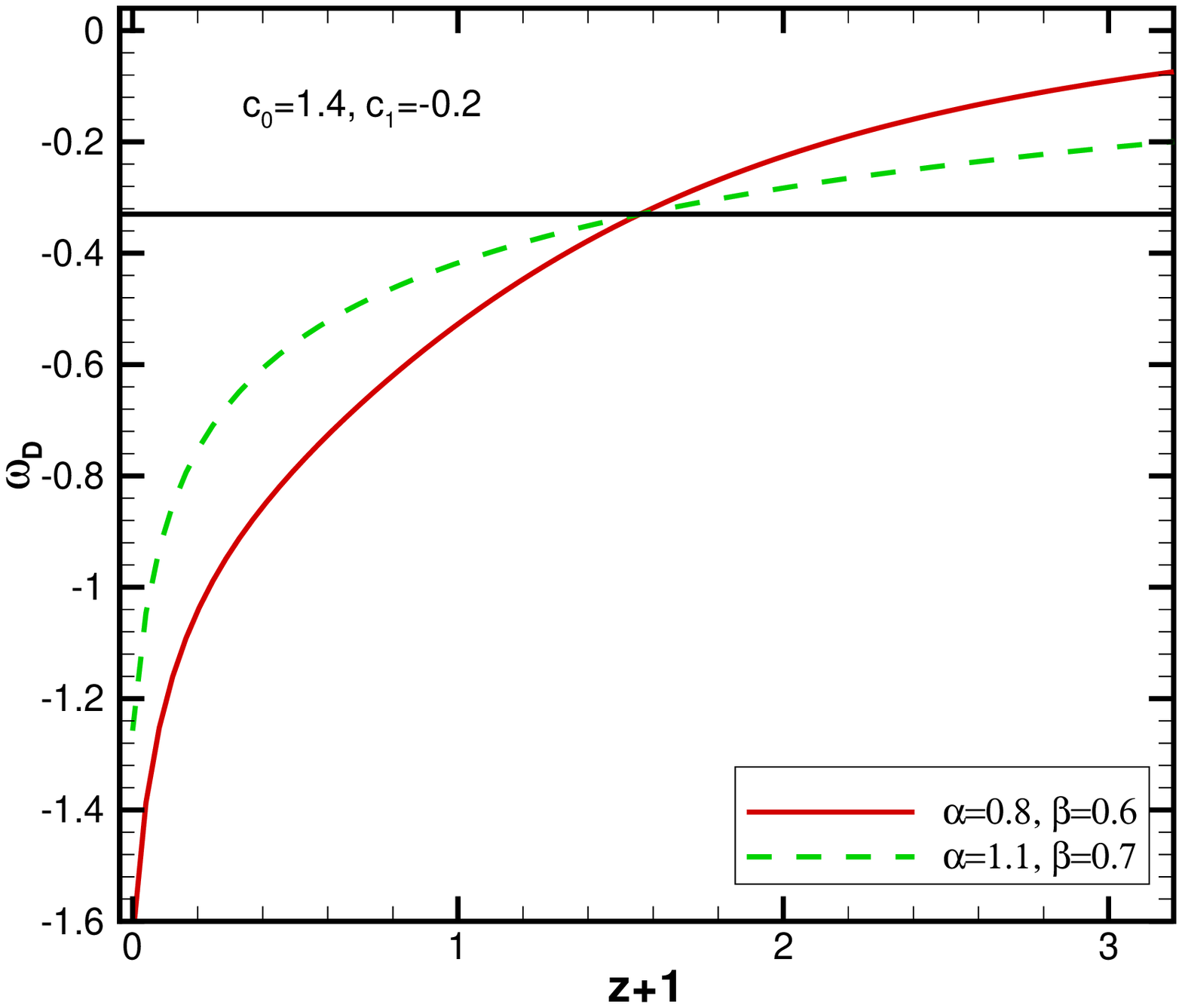}
\includegraphics[width=8cm]{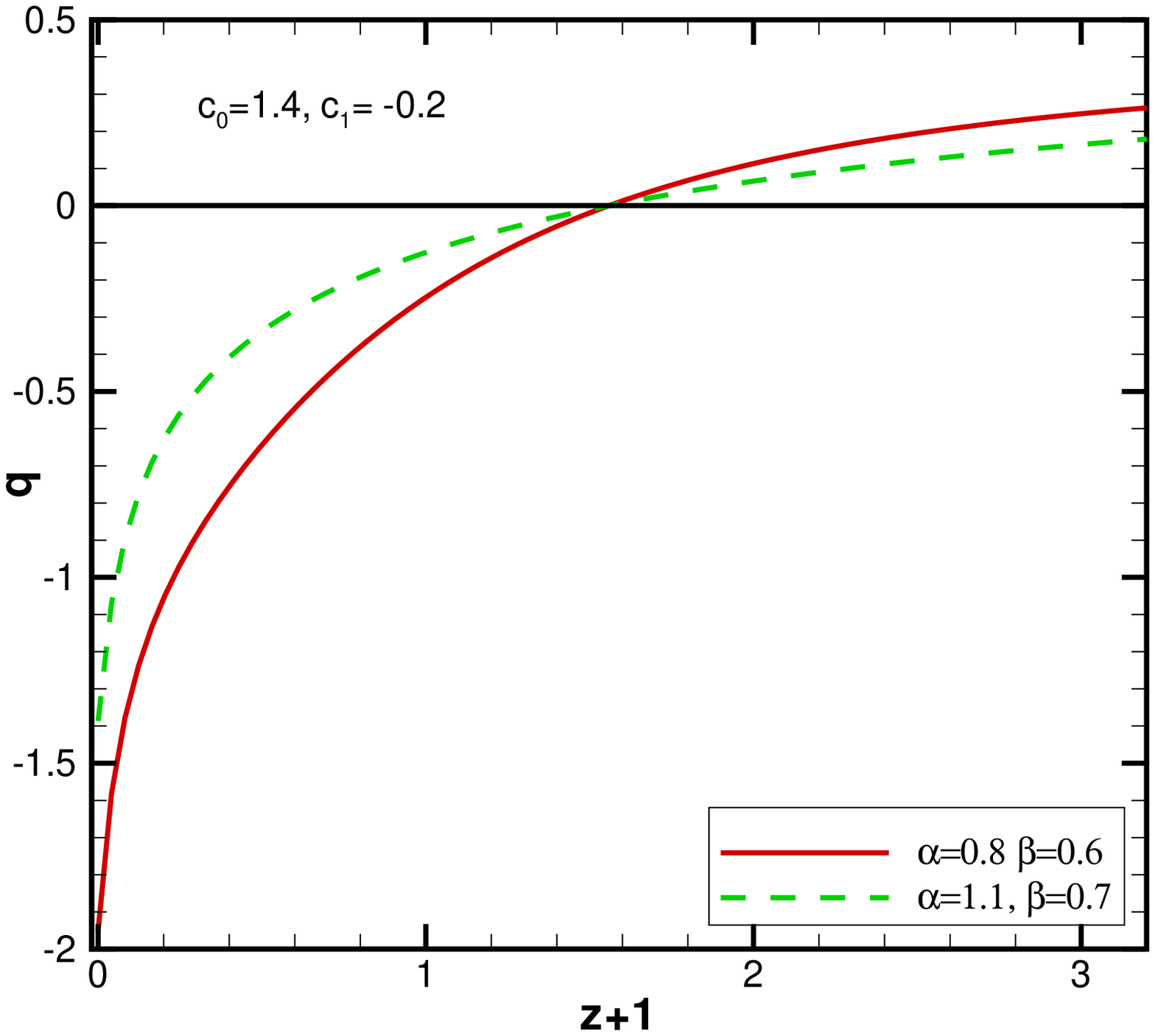}
\caption{The evolution of equation of state parameter $ \omega_D $
and deceleration parameter $ q $
        versus  $ 1+z $ for  GHDE3 model with GO cutoff. }\label{w3,q3-GO}
\end{center}
\end{figure}
\subsection{GHDE4: The Ma-Zhang type}
Using Eqs. (\ref{G4}) and (\ref{Omegaprime}), it is easy to show
that the evolution of the dimensionless GHDE4 density  can be
obtained as
\begin{equation}\label{Omegaprime4}
{\Omega}^{\prime}_D=(1-\Omega_D)\left(\frac{2\Omega_D(1+z)^2}{\beta[c_0(1+z)+c_1\ln(2+z)-c_1(1+z)\ln2]^2}-\frac{2\alpha}{\beta}+3\right).
\end{equation}
Fig. \ref{Omega3} shows  the evolution of the dimensionless GHDE
density parameter $ \Omega_D $. Again, at the early universe where
$ z\rightarrow \infty $ we have  $ \Omega_D\rightarrow 0$, while
at the late time where $ z\rightarrow -1 $, the DE dominated,
namely $ \Omega_D\rightarrow 1 $.
\begin{figure}[htp]
\begin{center}
\includegraphics[width=8cm]{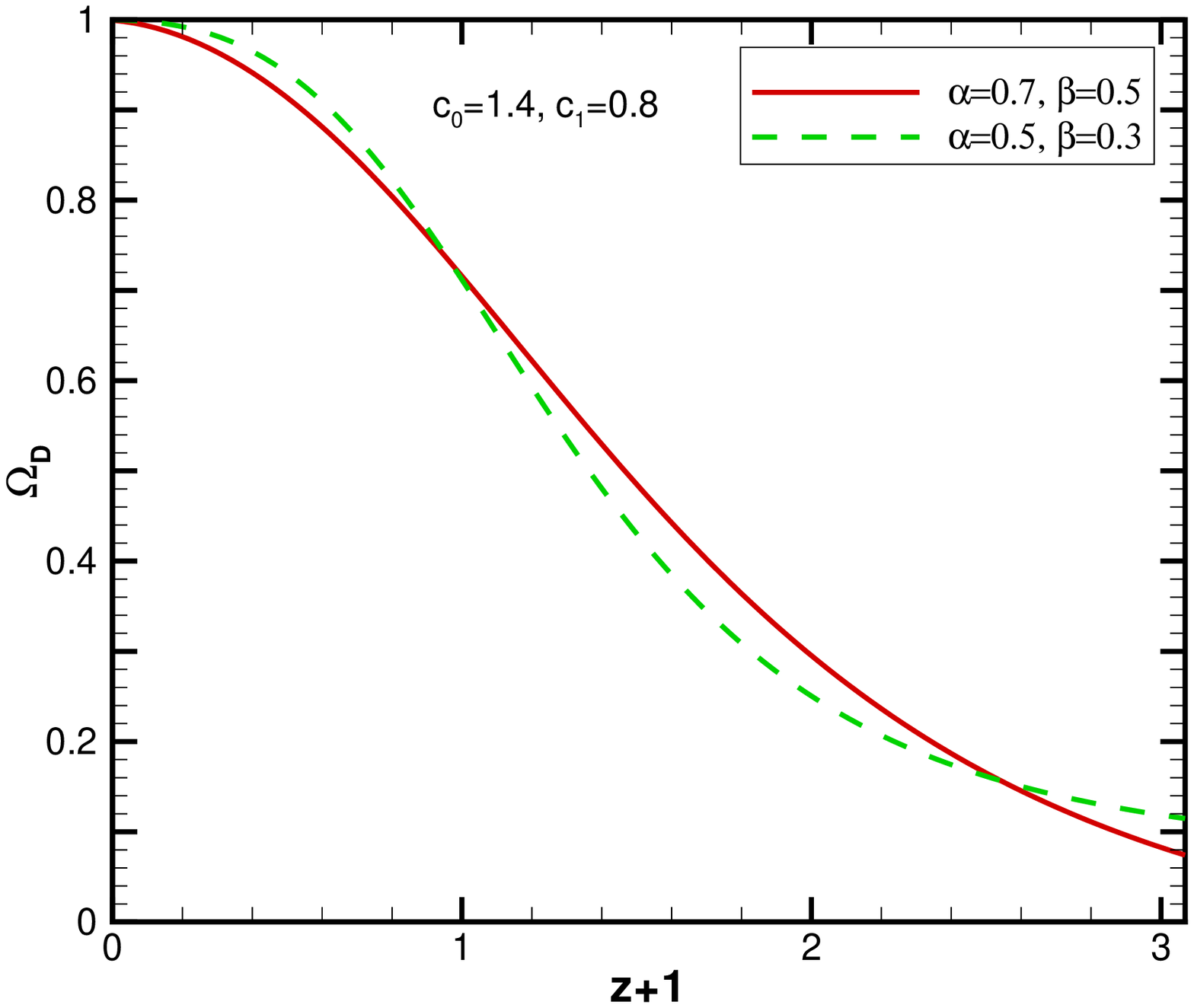}
\caption{The evolution of the dimensionless density parameter $
\Omega_D $
        versus $ 1+z $ for  GHDE4 model with GO cutoff}\label{Omega4}
\end{center}
\end{figure}
Inserting Eq. (\ref{G4}) in Eqs. (\ref{wD2}) and (\ref{q3}) we can
obtain the EoS and the deceleration parameters as follows
\begin{eqnarray}\label{wGO4}
\omega_D(z)&=&\frac{2\alpha}{3\beta
\Omega_D}-\frac{2(1+z)^2}{3\beta[c_0(1+z)+c_1\ln(2+z)-c_1(1+z)\ln
2]^2}-\frac{1}{\Omega_D},\\
q(z)&=&-1-\frac{\Omega_D(1+z)^2}{\beta[c_0(1+z)+c_1\ln(2+z)-c_1(1+z)\ln
2]^2}+\frac{\alpha}{\beta}. \label{qGO4}
\end{eqnarray}
The behavior of the EoS parameter $\omega_D(z)$ and the
deceleration parameter $ q(z)$ are plotted in Figs. \ref{w4,q4-GO}
and \ref{w5,q5-GO}. From Fig. \ref{w4,q4-GO}, we see that when we
fix $ 1\leq c_0\leq 2$ and $-1\leq c_1\leq 0 $ and vary the
parameters $0\leq\alpha,\beta\leq 1 $, the GHDE4 can behave as
phantom DE model at the future, while for $ c_0,c_1>0 $ and $
1<\alpha,\beta<2 $ the EoS parameter cannot cross the phantom line
and is always larger than $-1 $ (see Fig. \ref{w5,q5-GO}).
\begin{figure}[htp]
\begin{center}
\includegraphics[width=8cm]{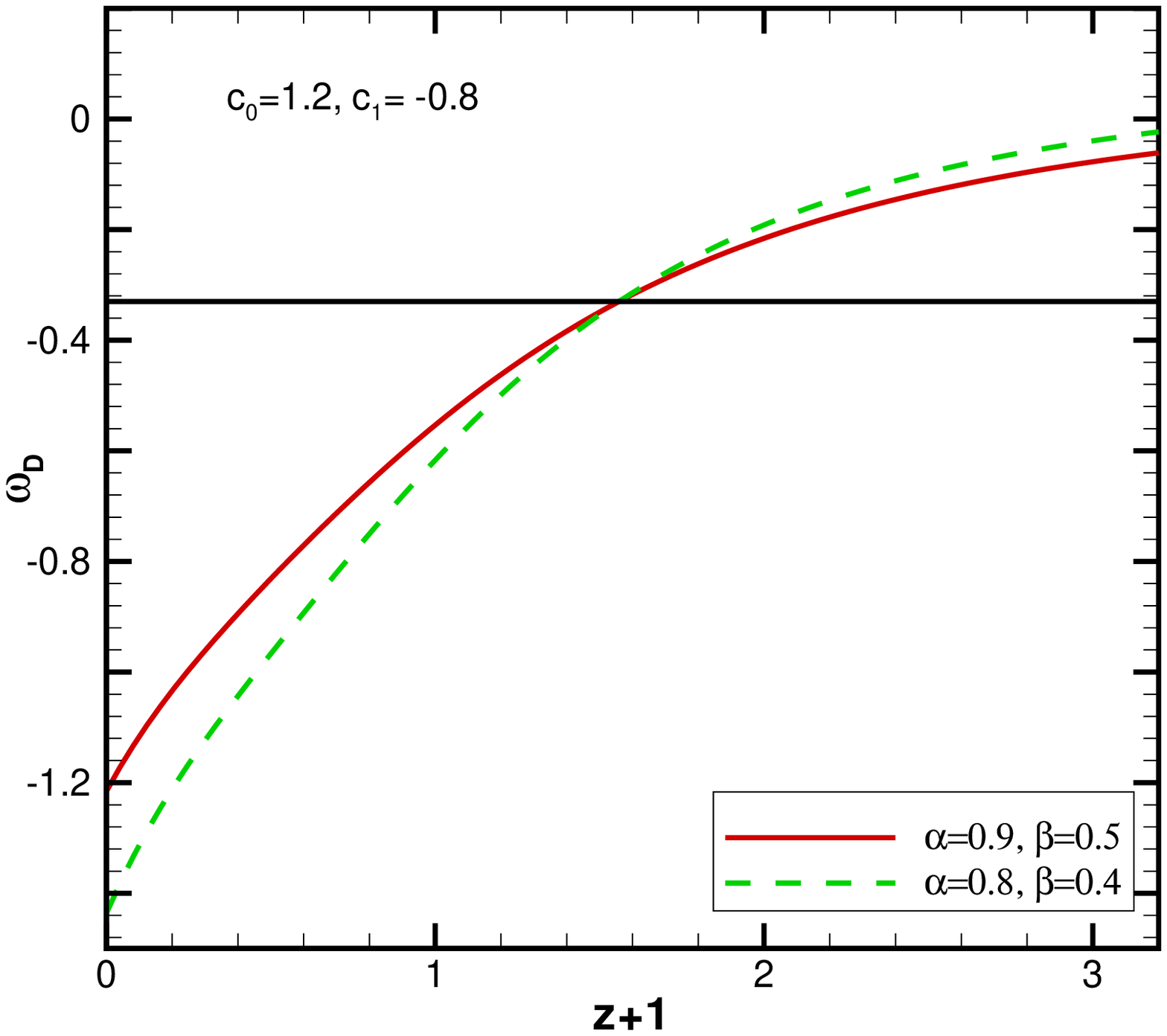}
\includegraphics[width=8cm]{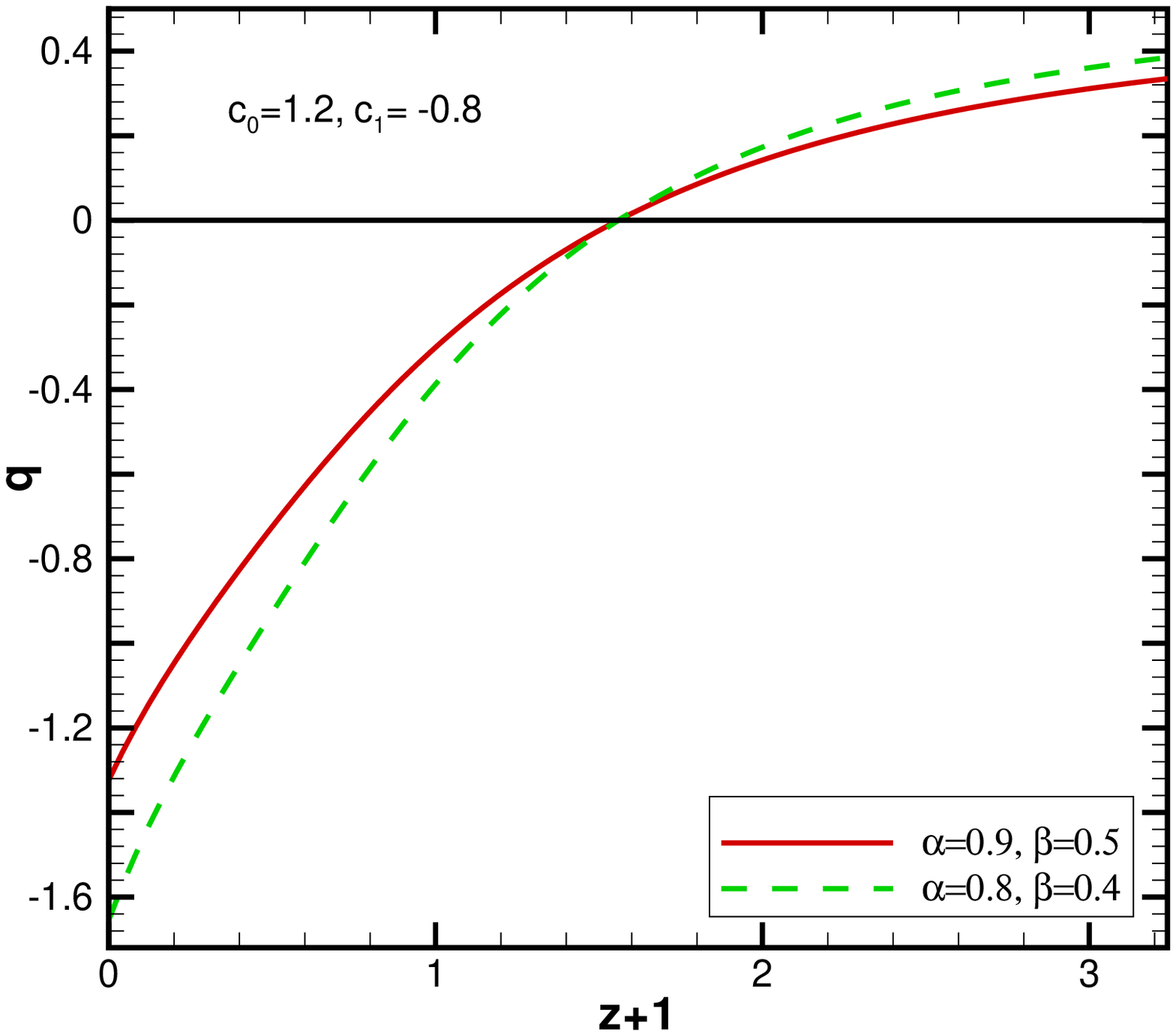}
\caption{The evolution of equation of state parameter $ \omega_D $
and deceleration parameter $ q $
        versus  $ 1+z $ for  GHDE4 model with GO cutoff. }\label{w4,q4-GO}
\end{center}
\end{figure}
\begin{figure}[htp]
\begin{center}
\includegraphics[width=8cm]{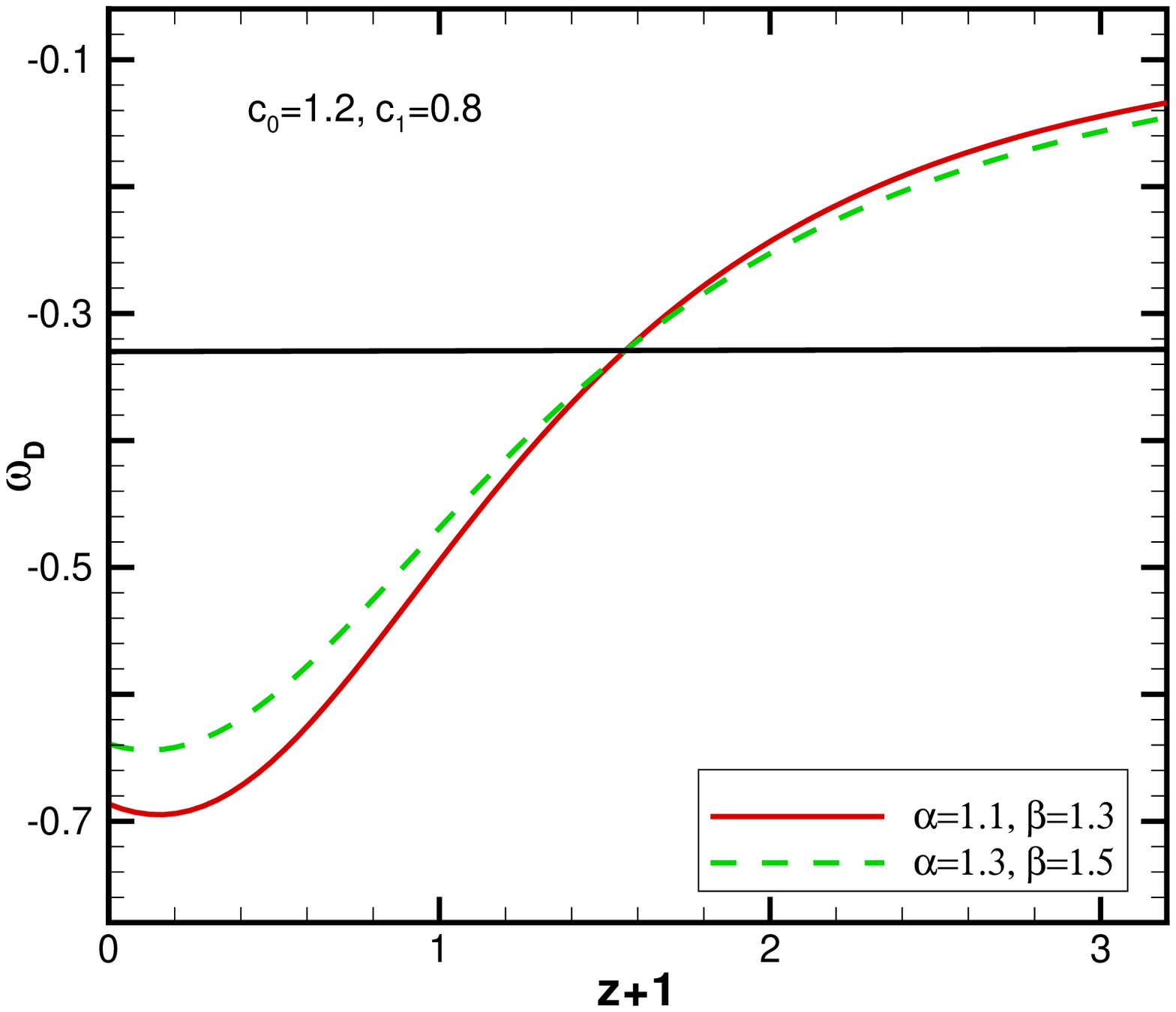}
\includegraphics[width=8cm]{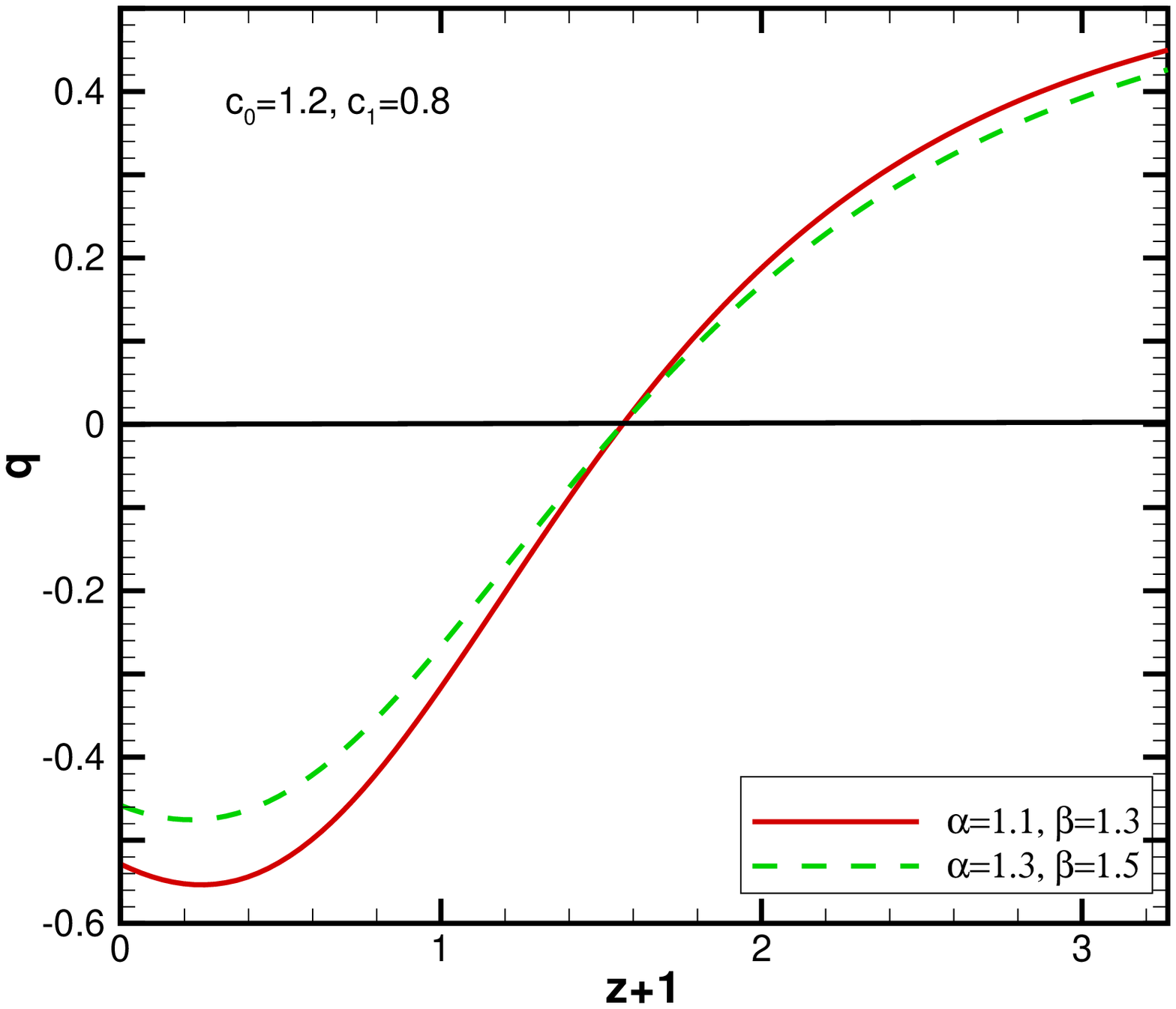}
\caption{The evolution of equation of state parameter $ \omega_D $
and deceleration parameter $ q $
        versus $ 1+z $ for  GHDE4 model with GO cutoff. }\label{w5,q5-GO}
\end{center}
\end{figure}
\section{Conclusion and discussion}
In this paper, we have studied HDE with time varying parameter $
c^2 $, the so-called generalized holographic dark energy (GHDE),
in a spatially flat universe. It is important to note that, for
the sake of simplicity, very often the $c^2$ parameter in the HDE
model is assumed constant. However, in general, it can be regarded
as a function of redshift parameter $z$ during the history of the
universe. By choosing four parameterizations of $c(z)$, including
the CPL type, JBP type, Wetterich type and Ma-Zhang type
parameterizations  for $c(z)$, we have investigated the effects of
varying $c^2(z)$ term on the cosmological evolutions of GHDE
model. As system's IR cutoff we have considered the Hubble radius
$ L=H^{-1} $ and the GO cutoff, $ L=(\alpha H^2+\beta
\dot{H})^{{-1}/{2}} $. We have investigated the evolution of EoS
and deceleration parameters for all these parameterizations. We
found that in all GHDE models, with both Hubble and GO cutoffs,
the universe has a transition from a deceleration to an
acceleration phase during its history. We have found that for
Hubble cutoff, the EoS parameter of GHDE can realize a quintom
behavior; namely, it evolves from a quintessence-like component to
a phantom-like component. While for the GO cutoff, not only at the
present time ($z=0$), but also at the far future where
$1+z\rightarrow0$, the EoS parameter, $\omega_D$, cannot cross the
phantom line and we have always $\omega_D>-1$ unless in a very
special case.

It is worth mentioning that the simple and natural choice for the
system's IR cutoff in the HDE model, is the Hubble radius $
L=H^{-1} $. However, it was argued that this choice for the IR
cutoff leads to a wrong equation od state for dark energy, namely
$\omega_D=0$ \cite{Hsu}, unless the interaction between two dark
components of the universe is taken into account \cite{Pavon}. In
this paper, we demonstrated that by taking into account the time
varying parameter $c(z)$  can leads to an accelerated universe for
$ L=H^{-1} $  IR cutoff, even in the absence of the interaction
between two dark components of energy of the universe. Besides, by
suitably choosing of the parameter, not only the accelerated
universe can be achieved, but also the EoS parameter can cross the
phantom line $\omega_D=-1$, even in the absence of interaction. As
far as we know, this is a new result, which has not been reported
already. In order to more investigate the behavior of the EoS and
deceleration parameters, we plotted the evolution of these
parameters versus redshift parameter $z$. From these figures we
see that has a decelerated phase at the early time $ (z\rightarrow
\infty) $ and encounters a phase transition to an accelerated
phase around  $z\approx0.6$ which is consistent with recent
observations \cite{Daly,Kom1,Kom2,Planck1,Planck2}.
\acknowledgments{We thank from the Research Council of Shiraz
University. This work has been supported financially by Research
Institute for Astronomy \& Astrophysics of Maragha (RIAAM), Iran.}

\end{document}